\documentclass[a4paper,11pt]{article}
\usepackage{pos}
\newcommand{\be}{\begin{equation}}
\newcommand{\ee}{\end{equation}}
\newcommand{\bea}{\begin{eqnarray}}
\newcommand{\eea}{\end{eqnarray}}

\newcommand{\arXiv}[2]{\href{http://arxiv.org/pdf/#1}{{\tt [#2/#1]}}}
\newcommand{\arXivold}[1]{\href{http://arxiv.org/pdf/hep-#1}{{\tt [#1]}}}
\def\bma#1{\mbox{\boldmath{$#1$}}}

\title{The Unreasonable Effectiveness of the Tunneling Potential}

\author{Jos\'e R. Espinosa}

\affiliation{Instituto de F\'{\i}sica Te\'orica, IFT-UAM/CSIC,
C/ Nicol\'as Cabrera 13-15, Campus de Cantoblanco, 28049, Madrid, Spain}

\emailAdd{jr.espinosa@csic.es}

\abstract{The Tunneling Potential Formalism was introduced to calculate the tunneling actions that control vacuum decay as an alternative to the standard Euclidean Formalism. The new approach sets the problem as a simple variational problem in field space with decay described by a tunneling potential function $V_t$ that extremizes a simple action functional $S[V_t]$ and has a number of appealing properties that have been presented elsewhere. In this note I discuss several instances in which this $V_t$ approach seems to give more than one would have expected a priori, as the following: the $V_t$ describing the decay is a minimum of the new action $S[V_t]$ rather than a saddle point; the decay of
AdS, dS or Minkowski vacua are governed by a unique universal $S[V_t]$
which also gives the Hawking-Moss instanton in the appropriate limit;
physically relevant solutions beyond the Coleman-De Luccia (CdL) bounce, like pseudo-bounces or bubbles of nothing (BoNs), show up in a straightforward way as generalizations of the CdL bounce, with the correct boundary conditions; in cases for which the Euclidean action calculation requires the inclusion of particular boundary terms (like for BoNs or for the decay of AdS maxima above the Breitenlohner-Freedman bound) $S[V_t]$ gives the correct result without the need of including any boundary term.
}

\FullConference{Corfu Summer Institute 2023 "School and Workshops on Elementary Particle Physics and Gravity" (CORFU2023)\\
 23 April - 6 May , and 27 August - 1 October, 2023\\
Corfu, Greece\\}

\begin{document}
\maketitle

\section{Introduction}

In his famous article \cite{Wigner}, "The Unreasonable Effectiveness of Mathematics in the Natural Sciences", published in 1960, Eugene Wigner discussed, among other things, how mathematics is unreasonably useful in the natural sciences, how mathematical concepts turn up in unexpected connections and how this prevent us from knowing if a theory formulated in some particular  mathematical form is uniquely appropriate to describe nature.

In the same spirit (if not the scope!) of that work, in these proceedings I  discuss some remarkable properties of the so-called tunneling potential formalism, which has been proposed as an alternative formulation of the problem of vacuum decay in quantum field theory. In doing so, I depart from my actual talk at the workshop. As the material actually covered in the talk can be found already in the original research articles (see list of references) I take the opportunity to present this discussion, which would not fit in a regular journal article. 

After introducing the formalism in the next section, both with and without gravitational corrections, I will detail in the following sections some of the good properties of the approach, distinguishing between those that could be anticipated from those that come as unexpected (and welcome) surprises.

\section{The Tunneling Potential Formalism}
The calculation of tunneling actions (which control the exponential suppression of vacuum decay in quantum field theory) is routinely performed using the powerful and elegant method developed by Coleman \cite{Coleman}. Briefly, one looks for an Euclidean bounce configuration that connects the initial state (the homogeneous false vacuum) at Euclidean time $\tau=-\infty$ to the final state of the tunneling process (the nucleated bubble of the deeper phase) at $\tau=0$ and bounces back to the initial configuration at $\tau=\infty$. The Euclidean action associated to such bounce is the action that suppresses the vacuum decay rate and the bounce is a solution of the Euler-Lagrange equation associated to extremizing such action. The task of finding the bounce is enormously simplified by the fact that it is invariant under rotations in the Euclidean spacetime and thus one simply has to determine a single function $\phi_B(r)$ describing the bounce profile (with $r$ a radial coordinate, with $r=\infty$ corresponding to the false vacuum and $r=0$ probing the deeper part of the potential) and solving
\be
\ddot\phi+\frac{3\dot\phi}{r}=V'\ ,
\label{EoMphi}
\ee
where $V(\phi)$ is the potential of the scalar field, $\dot x\equiv dx/dr$ and $x'\equiv dx/d\phi$. The boundary conditions are
\be
\dot\phi(0)=0\ ,\quad\quad \phi(\infty)=\phi_+\ ,\quad\quad \phi'(\infty)=0\ ,\label{BCs}
\ee
where $\phi_+$ is the false vacuum and $\phi_0\equiv \phi(0)$ has to be found so as to satisfy these boundary conditions. Once the bounce $\phi_B(r)$ is found, the tunneling exponent in the decay rate per unit volume ($\Gamma/V\sim e^{-S}$) is given by the difference of the Euclidean actions for bounce and false vacuum background, $S=\Delta S_E\equiv S_E[\phi_B]-S_E[\phi_+]$.

In a seminal paper, Coleman and De Luccia \cite{CdL} discussed the impact of gravitational effects on the process of vacuum decay. 
When gravity is taken into account \cite{CdL}, one needs to assume rotational symmetry (a reasonable but unproven point) and include the spacetime metric to describe the gravitational effects of the bounce. The line-element can be written as
\be
ds^2=dr^2+\rho(r)^2d\Omega_3^2\ ,
\label{CdLmetric}
\ee
where $d\Omega_n^2$ is the line element on the n-sphere of unit radius. Therefore, one has to find the two functions $\phi_B(r)$ (the bounce) and $\rho_B(r)$ (the metric function) which are solutions of the Euler-Lagrange equations for the gravitationallly corrected action, the system of coupled differential equations
\bea
&&\ddot\phi+\frac{3\dot\rho\dot\phi}{\rho}=V'\ ,
\label{EoMphig}\\
&& \dot\rho^2 =1+ \frac{\kappa}{3}\rho^2\left(\frac12\dot\phi^2-V\right)\ ,
\label{EoMrho}
\eea
where $\kappa=1/m_P^2$ and $m_P$ is the reduced Planck mass. The $r=0$ boundary conditions are now
\be
\phi(0)=\phi_0\ ,\quad \dot\phi(0)=0\ ,\quad \rho(0)=0\ .
\ee
For the decay of AdS or Minkowski vacua the boundary conditions for the field at $r=\infty$ are as in (\ref{BCs}) and $\rho(\infty)=\infty$.
For the decay of dS vacua the range of $r$ is finite, $r\in (0,r_{max})$,
with $\rho(r_{max})=0$ and the field does not reach the false vacuum value,
\be
\phi(r_{max})\equiv\phi_{0+}\neq \phi_+\ , \quad \dot\phi(r_{max})=0\ .
\ee
Once the bounce $\phi_B(r)$  and its associated metric function $\rho_B(r)$ are found, the tunneling action is the difference of the Euclidean actions for bounce and false vacuum background, $S=\Delta S_E\equiv S_E[\phi_B,\rho_B]-S_E[\phi_+,\rho_+]$, where $\rho_+$ is the metric function for the background.

Besides modifying quantitatively the vacuum decay rates, gravitational effects could lead to qualitative changes: gravity could prevent AdS or Minkowski false vacuum decay (gravitational quenching) and dS vacua could tunnel upwards to other dS vacua with higher potential energy.

\subsection{$\bma{V_t}$ formalism without gravity}
\begin{figure}[t!]
\includegraphics[width=0.46\textwidth]{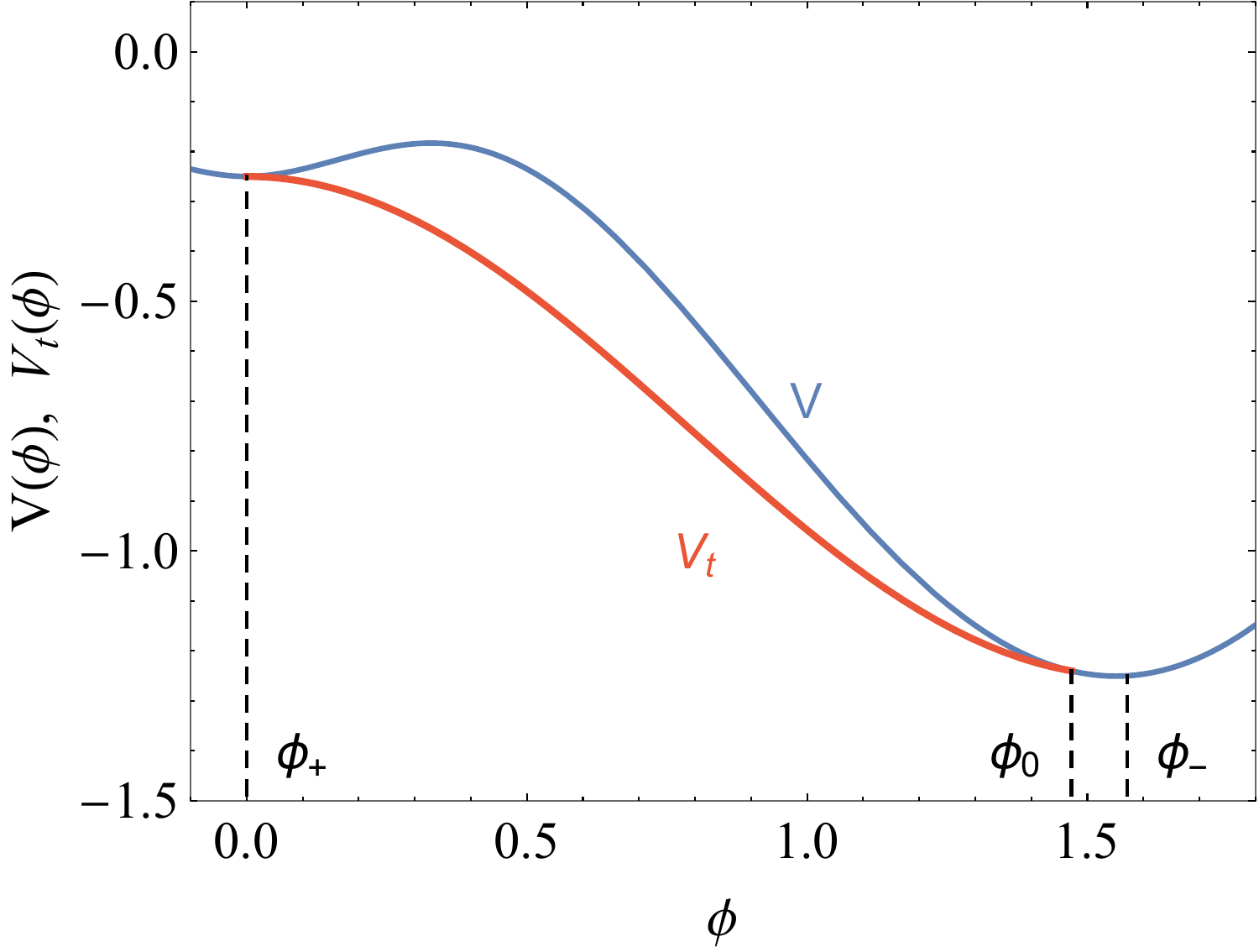}\hspace{0.3cm}
\includegraphics[width=0.46\textwidth]{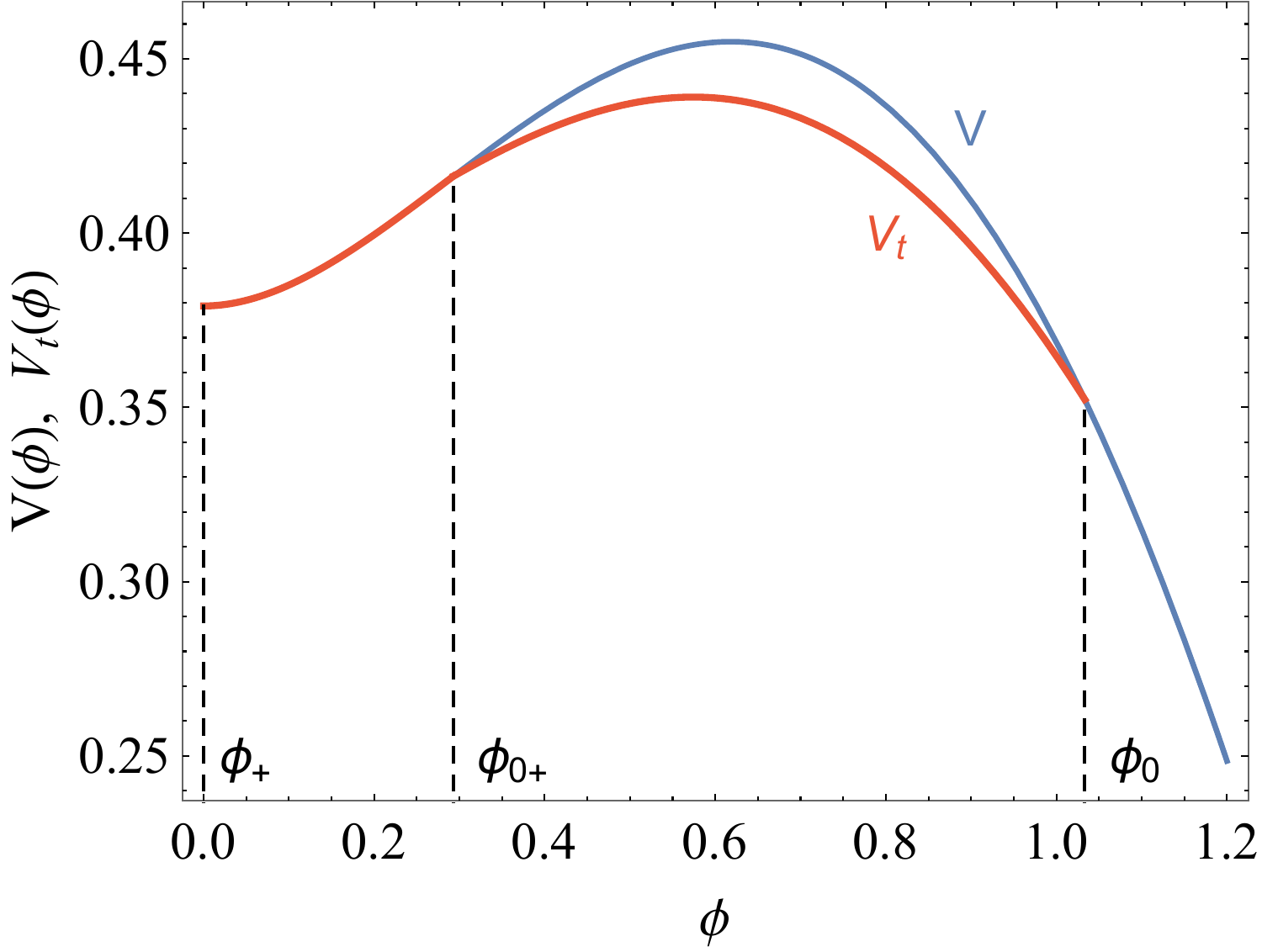}
\caption{\em Tunneling potential $V_t(\phi)$ (red) for decay out of an
AdS vacuum (upper plot) or a dS vacuum (lower plot) in some example potentials $V(\phi)$ (blue).} 
\label{fig:Vt}
\end{figure}

The Euclidean approach just described is not necessary to calculate the tunneling action and \cite{E} showed how one can get rid of all Euclidean quantities to transform the problem into a variational problem in field space. In this new formulation one considers possible decay paths out of the false vacuum defined by some function $V_t(\phi)$ called tunneling potential, with $V_t\leq V$, for an example see Fig.~\ref{fig:Vt}, left plot. To any such path one associates a number given by the functional
\be
S[V_t]=54\pi^2\int_{\phi_+}^{\phi_0}\frac{(V-V_t)^2}{-V_t'{}^3}d\phi\ ,
\label{SVt}
\ee
where $\phi_+$ is the false vacuum and $\phi_0$ is the field value where the
decay path ends (by convention $\phi_+<\phi_0$). Then one searches for the path that minimizes this functional and the minimal value of $S[V_t]$ gives precisely the tunneling action for the decay, that is, $S[V_t]=\Delta S_E$. Moreover, the end point value $\phi_0$ coincides with the value of the Euclidean bounce at its center, $\phi_0=\phi_B(0)$.

How is (\ref{SVt}) obtained? The original derivation in \cite{E} took the Euclidean bounce as starting point and got rid of all Euclidean quantities by using the formula 
\be
V_t(\phi)=V(\phi)-\frac12 \dot\phi_B^2(\phi)\ .
\ee
Here, $\dot\phi_B(\phi)$ is the bounce slope $d\phi_B/dr$ expressed as a function of $\phi$ by solving $\phi_B(r)=\phi$ for $r(\phi)$  [which can be done as $\phi_B(r)$ is monotonic]. Notice that this is {\it not} the definition of $V_t$ but just the key link between the two formalisms.
From this formula and its derivatives one gets
\be
\dot\phi=-\sqrt{2(V-V_t)}\ ,\quad \ddot\phi=V'-V_t'\ ,
\ee
and from the bounce equation (\ref{EoMphi}) one also gets
\be
r=\frac{3\sqrt{2(V-V_t)}}{-V_t'}\ .
\label{r}
\ee
Using these formulas (whose left hand sides are Euclidean quantities associated to the bounce while the right hand sides contain only the field-dependent quantities $V$ and $V_t$) one can transform the Euclidean bounce variational problem and its action into a variational problem for $V_t$ with action (\ref{SVt}). 

Besides this direct derivation, one can also arrive at $S[V_t]$ in an straightforward way by using a canonical transformation of the Euclidean formulation \cite{EJK}. Finally, the simplest way of arriving at $S[V_t]$ is to consider as in \cite{Ethin} the general Euclidean bounce as an infinite collection of infinitesimal concentric shells to which a thin-wall action can be assigned. The integral over all these shells reproduces the tunneling action  (\ref{SVt}).

It is not surprising that one can reformulate the calculation of the tunneling action in field space as explained. However, the new formalism has an unexpected property: while the Euclidean bounce is a saddle-point of the Euclidean action functional, the tunneling potential associated to vacuum decay is a minimum of $S[V_t]$ (for a proof, see appendix of \cite{E}). This is of course welcome for numerical applications (e.g. for multifield decay problems, see \cite{EK}) as it is an easier task to look for a minimum rather than a saddle point. Nevertheless, the reformulation was not built with this aim in mind and there was no reason to expect this to happen.

The tunneling potential has other interesting properties (e.g. it gives a very intuitive picture of the tunneling process; $V_t$ is a monotonically decreasing function and this makes it easy to approximate numerically \cite{E}; it is ideally suited to discuss vacuum decay when there is no bounce \cite{ENoB}; etc.) but these are perhaps less striking and fall into the potential advantages one would expect of different formulations of a problem.

\subsection{$\bma{V_t}$ formalism with gravity\label{Vtgrav}}
After getting rid of Euclidean quantities as discussed in the previous subsection, the natural question to ask was whether this would still be possible (or useful) once gravitational effects were included. The question was answered in the positive by \cite{Eg}: the tunneling action for decay can be calculated in exactly the same manner but with a different action functional which now takes the form 
\be
S[V_t]=\frac{6\pi^2}{\kappa^2}\int_{\phi_+}^{\phi_0}\frac{(D+V_t')^2}{D V_t^2}d\phi\ ,
\label{SVtg}
\ee
where
\be
D\equiv \sqrt{V_t'{}^2+6\kappa(V-V_t)V_t}\ .
\ee
The methods one can use to get this action are the same that were mentioned in the previous section for the case without gravity. In this case we
still have $\dot\phi=-\sqrt{2(V-V_t)}$ and (\ref{r}) is replaced by
\be
\rho=\frac{3\sqrt{2(V-V_t)}}{D}\ .
\label{rho}
\ee
Now, the fact that this is possible to achieve, so that the tunneling is described by a single function $V_t(\phi)$ instead of the two functions [$\phi_B(r)$ and $\rho_B(r)$] necessary in the Euclidean formulation, is a nice property of the $V_t$ approach but, in hindsight, is not surprising either. After all, $\rho_B(r)$ is obtained from the differential equation
(\ref{EoMrho}) which enforces a constraint. One could rewrite the field profile as a function of $\rho$, taken as an independent variable, and reduce the system of coupled equations (\ref{EoMphig}) and (\ref{EoMrho}) to a single equation
\be
\left(1-\frac{\kappa}{3}\rho^2V\right)\left(\phi''+\frac{3}{\rho}\phi'-\frac{\kappa}{3}\rho\phi'{}^3\right)=\left(V'+\frac{\kappa}{3}\rho V \phi'\right)\left(1-\frac{\kappa}{6}\rho^2\phi'{}^2\right)\ ,
\ee
where now $\phi'\equiv d\phi(\rho)/d\rho$. Even though this equation is more complicated, it serves the purpose of showing that the problem is really a single-field problem.\footnote{If the profiles $\phi(r)$ and $\rho(r)$ are needed, they can be obtained from $\phi(\rho)$ via $\dot\rho^2=(1-\kappa \rho^2 V/3)/(1-\kappa \rho^2\phi'{}^2/6)$.\label{foot}}

In the $V_t$ formalism, the shape of $V_t$ is qualitatively different depending on the potential energy density at the false vacuum, $V_+\equiv V(\phi_+)$. For AdS or Minkowski false vacua, $V_t$ is monotonically decreasing as in the absence of gravity. This case is illustrated in the left plot of Fig.~\ref{fig:Vt}. If the decaying vacuum is dS, there are two regimes in the $V_t$ solution:
From the false vacuum at $\phi_+$ to some value $\phi_{0+}$ in the increasing slope of $V$, one has $V_t\equiv V$ and, from $\phi_{0+}$ to some $\phi_0$ beyond the top of the barrier, $V_t<V$, with $V_t$ having first a positive slope, reaching a maximum and then decreasing towards $\phi_0$, see right plot of Fig.~\ref{fig:Vt}. This second part of $V_t$ corresponds to the field interval covered by the Euclidean bounce solution. Notice that, in the first interval the link $V_t=V-\dot\phi^2/2$ is not really applicable as $\dot\phi$ is meaningless there, so that the derivation of (\ref{SVtg}), which was based on that relation, must be carefully take this into account for the dS case.  

\begin{figure}[t!]
\begin{center}
\includegraphics[width=0.46\textwidth]{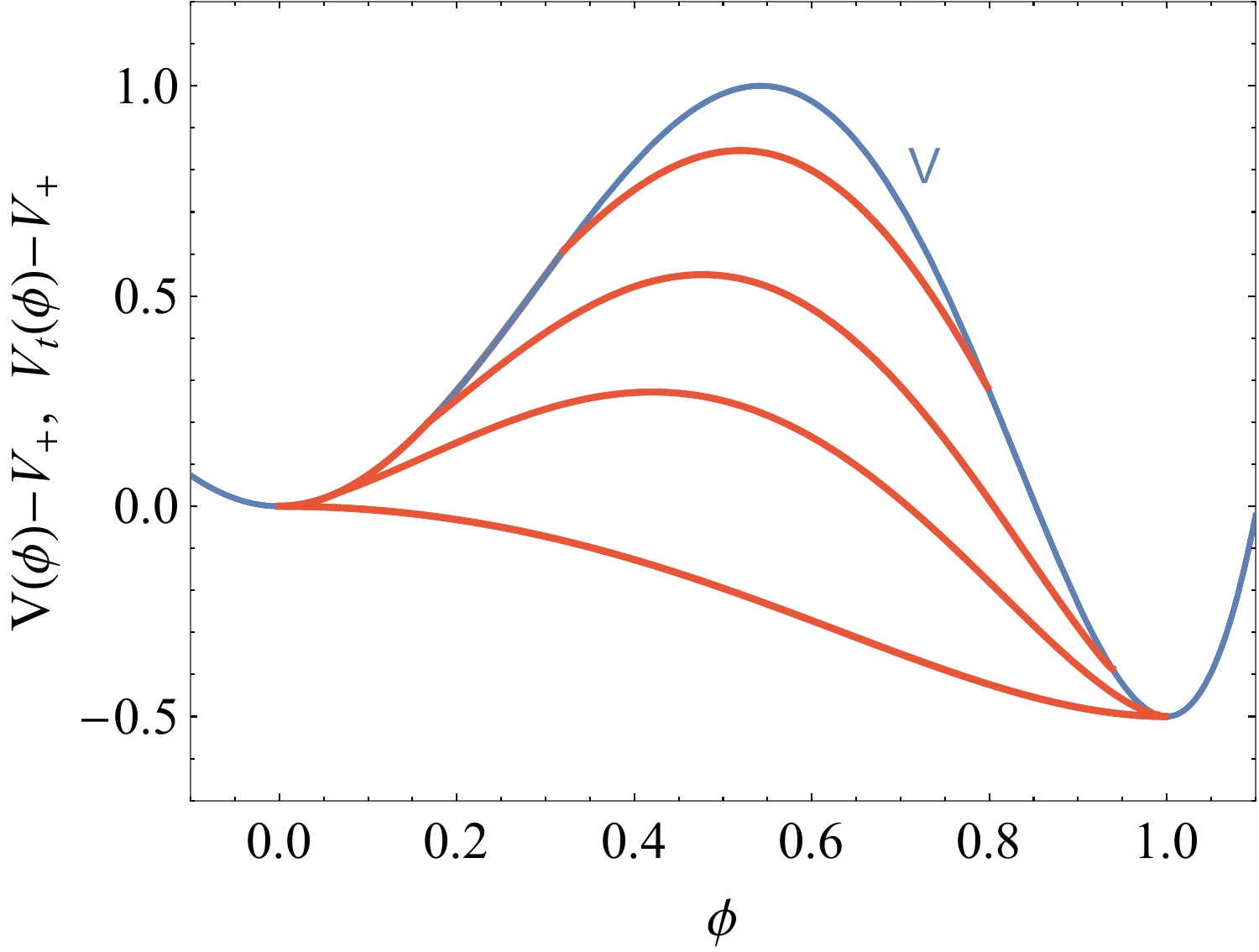}
\end{center}
\caption{\em 
Evolution of $V_t$ for a potential $V$ of fixed shape but with growing values of $V_+$ (both $V$ and $V_t$ are plotted subtracting $V_+$). The larger $V_+$ is, the higher $V_t$ lies and the smaller the CdL range is. Eventually, the CdL range disappears and only the Hawking-Moss solution remains, see text. 
} 
\label{fig:shrinkCdL}
\end{figure}

Remarkably, the tunneling action (\ref{SVtg}) gives the correct result\footnote{The general proof of $S[V_t]=\Delta S_E$ is in \cite{Eg} (see \cite{EF} for the proof in general spacetime dimension).} for any kind of vacuum without the need of treating the dS case separately. Moreover, when $V_+>0$ grows large enough, eventually the CdL interval shrinks to zero and there is no CdL bounce, see Fig.~\ref{fig:shrinkCdL}. In that case the vacuum decay still proceeds, but via the Hawking-Moss instanton \cite{HM}, with tunneling action
\be
S_{HM}=\frac{24\pi^2}{\kappa^2}\left(\frac{1}{V_+}-\frac{1}{V_T}\right)\ ,
\label{SHM}
\ee
where $V_T\equiv V(\phi_T)$ is the value of the potential at the top of the barrier, $\phi_T$. In the $V_t$ formalism, this action is recovered trivially: for sufficiently large $V_+$ the action is minimized by a configuration without CdL range, one has $V_t\equiv V$ from $\phi_+$ to $\phi_T$ and plugging this in (\ref{SVtg}) one gets precisely (\ref{SHM}). 

This "universal" validity of $S[V_t]$ in (\ref{SVtg}) for all types of vacua and capturing both CdL and Hawking-Moss results is another example of an unexpectedly simple result.  Things could have easily been more complicated as the derivation of $S[V_t]$ was based on the CdL bounce to start with and there was no reason to expect that the dS case, including the Hawking-Moss instanton, would fit into precisely the same general expression for $S[V_t]$.  

Besides this, the $V_t$ formalism has other nice (but more standard) properties to deal with gravitational corrections to false vacuum decay. For instance, it leads to a very direct understanding of gravitational quenching of the decay: for AdS or Minkowski vacua (which require $V_t<0$) it is not guaranteed that $D$ is real. In fact, if $D^2<0$ for any possible $V_t$ out of the false vacuum, then the decay is forbidden. For a given potential, this can be diagnosed by looking at the solution of $D=0$ that leaves out of the false vacuum [thus with boundary condition $\overline V_t(\phi_+)=V_+$], call it $\overline V_t$. By definition of $\overline V_t$, any $V_t$ with $D^2>0$ has a slope steeper than that of $\overline V_t$ and 
thus $V_t<\overline V_t$. If the potential is such that $\overline V_t$ does not reach $V$ after leaving the false vacuum, then no $V_t$ exists that can mediate vacuum decay.\footnote{In the critical case in which $\overline V_t$ reaches exactly the true vacuum, $\overline V_t$ is a $D=0$ configuration describing a flat domain wall between both vacua. Potentials that admit such domain wall configurations are of the general form $V=V_t-V_t'{}^2/(6\kappa V_t)$, as can be immediately derived from $D=0$. This reproduces the old result of \cite{Vc}.} For further details, see \cite{Eg,Eg2}.

\begin{figure}[t!]
\begin{center}
\includegraphics[width=0.46\textwidth]{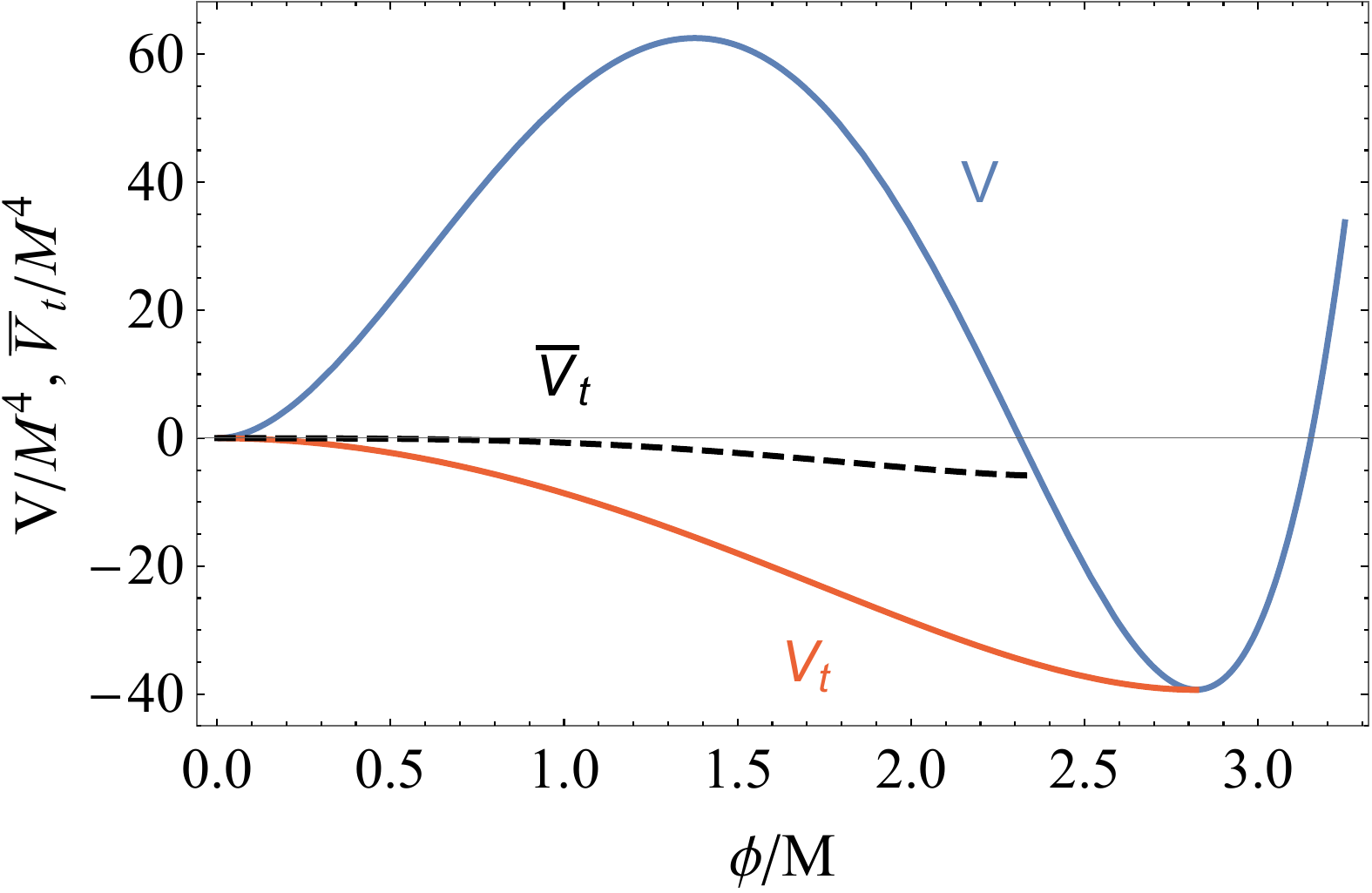}
\includegraphics[width=0.46\textwidth]{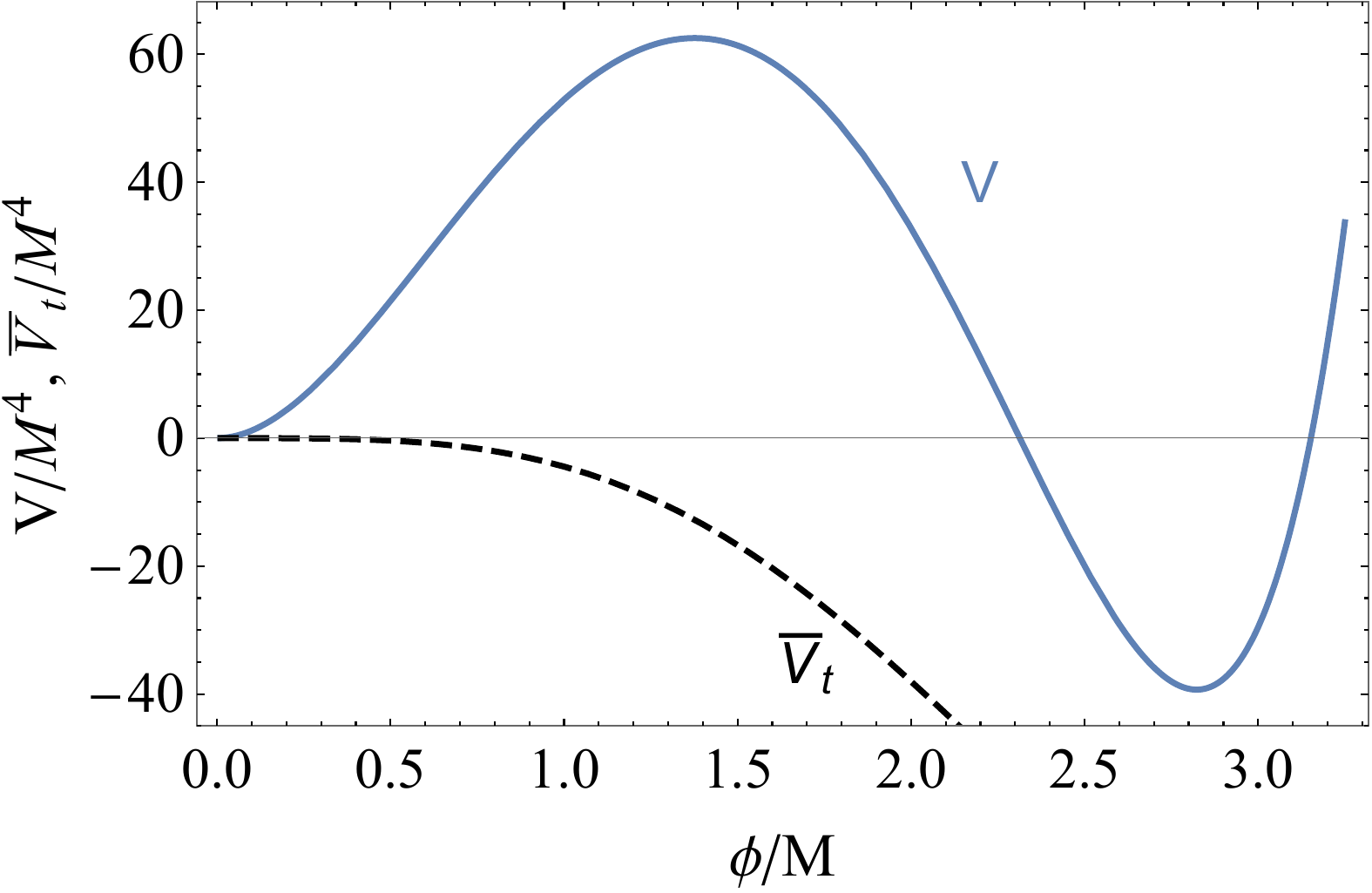}
\end{center}
\caption{\em 
Potential with a Minkowski false vacuum at $\phi_+=0$  for different strengths of gravitational effects (measured by $\kappa M^2$, with $M$ a typical mass scale of $V$). Left plot: subcritical case, with $\kappa M^2=0.02$; Right plot: supercritical case with $\kappa M^2=0.15$. The black dashed line gives the $D=0$, $\overline{V_t}$ solution. In the left plot the tunneling potential for vacuum decay is given by the red line, in the right no CdL decay is possible.
} 
\label{fig:subsuperV}
\end{figure}

\section{Pseudo-bounces}

To find numerical solutions for the $V_t$ function describing vacuum decay one needs to solve the corresponding Euler-Lagrange equation from the action (\ref{SVtg}), which reads
\be
(4V_t'-3V')V_t'+6(V-V_t)[V_t''+\kappa(3V-2V_t)]=0\ .
\label{EoMVt}
\ee
For AdS or Minkowski false vacua, the boundary conditions at the false vacuum are
\be
V_t(\phi_+)=V_+\ ,\quad V_t'(\phi_+)=0\ .
\label{BCMinkphip}
\ee
At the end point $\phi_0$ of the tunneling interval one has instead
\be
V_t(\phi_0)=V(\phi_0)\ , \quad V_t'(\phi_0)=\frac34 V'(\phi_0)\ .
\label{BCsphi0}
\ee
For dS vacua, the initial point of the tunneling is not the false vacuum but some higher value $\phi_{0+}$, and the boundary conditions are
\be
V_t(\phi_{0+})=V(\phi_{0+})\ ,\quad V_t'(\phi_{0+})=\frac34 V'(\phi_{0+})\ ,
\label{BCdSphi0p}
\ee
while for the end-point $\phi_0$ the boundary conditions are as in (\ref{BCsphi0}).

\begin{figure}[t!]
\begin{center}
\includegraphics[width=0.46\textwidth]{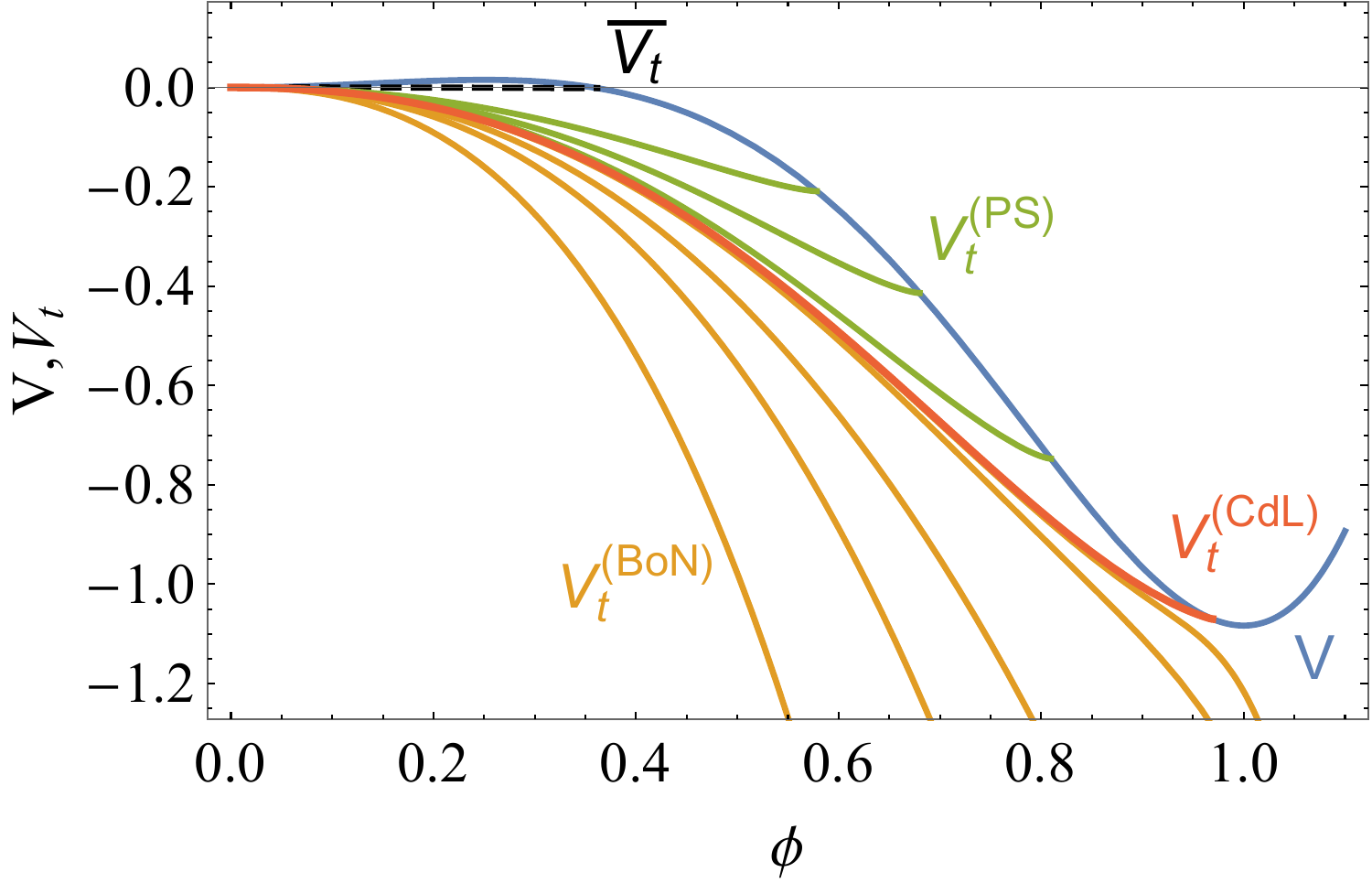}
\end{center}
\caption{\em 
Potential with a Minkowski false vacuum and tunneling potentials $V_t(A;\phi)$ of different types: $\overline{V_t}$ for $A=0$ (black dashed); pseudo-bounces for $0<A<A_{CdL}$ (green); CdL bounce for $A=A_{CdL}$ (red) and BoNs for $A>A_{CdL}$ (orange). 
} 
\label{fig:Minklowshoot}
\end{figure}

\begin{figure}[t!]
\includegraphics[width=0.46\textwidth]{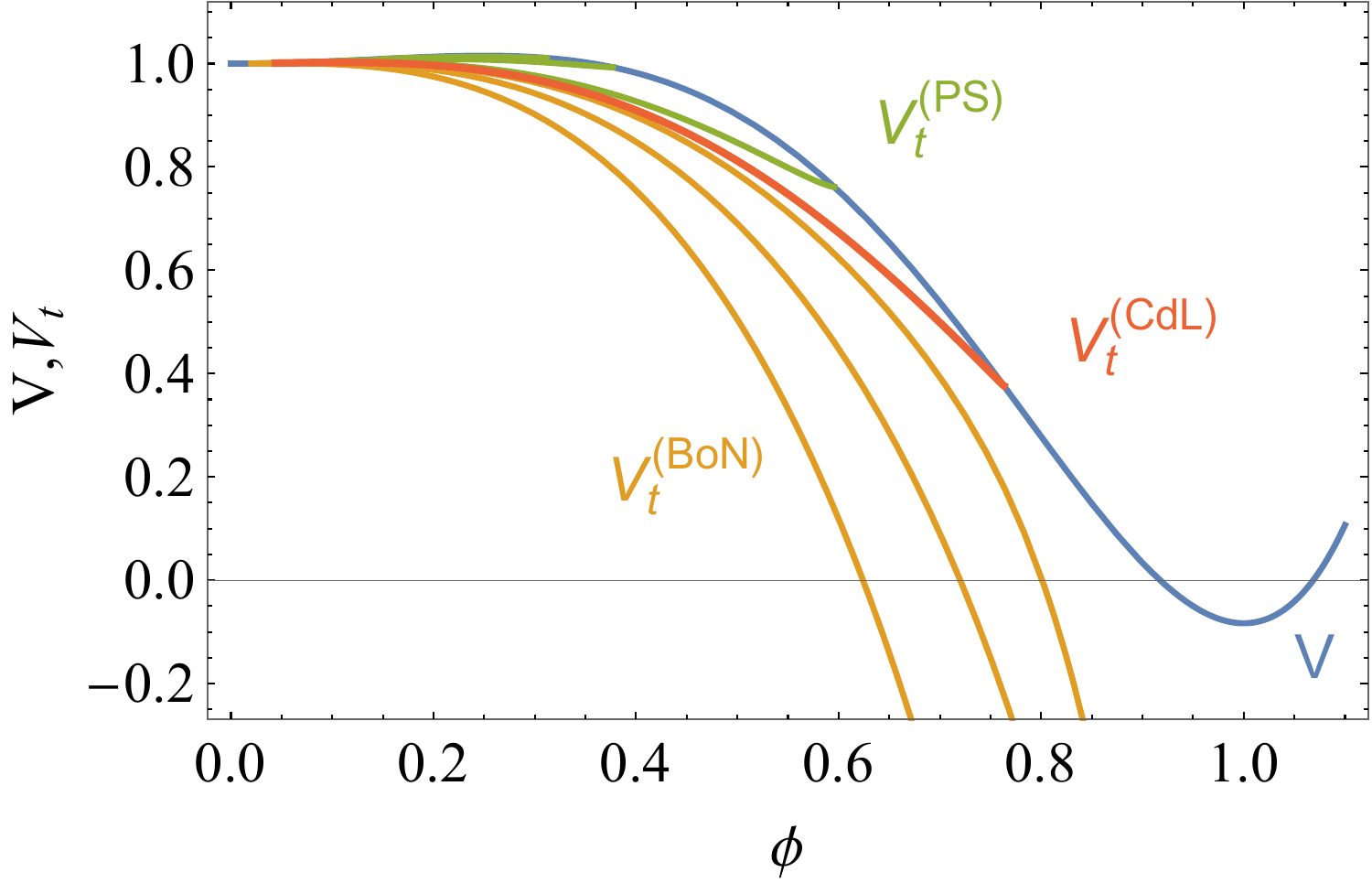}\hspace{0.3cm}
\includegraphics[width=0.46\textwidth]{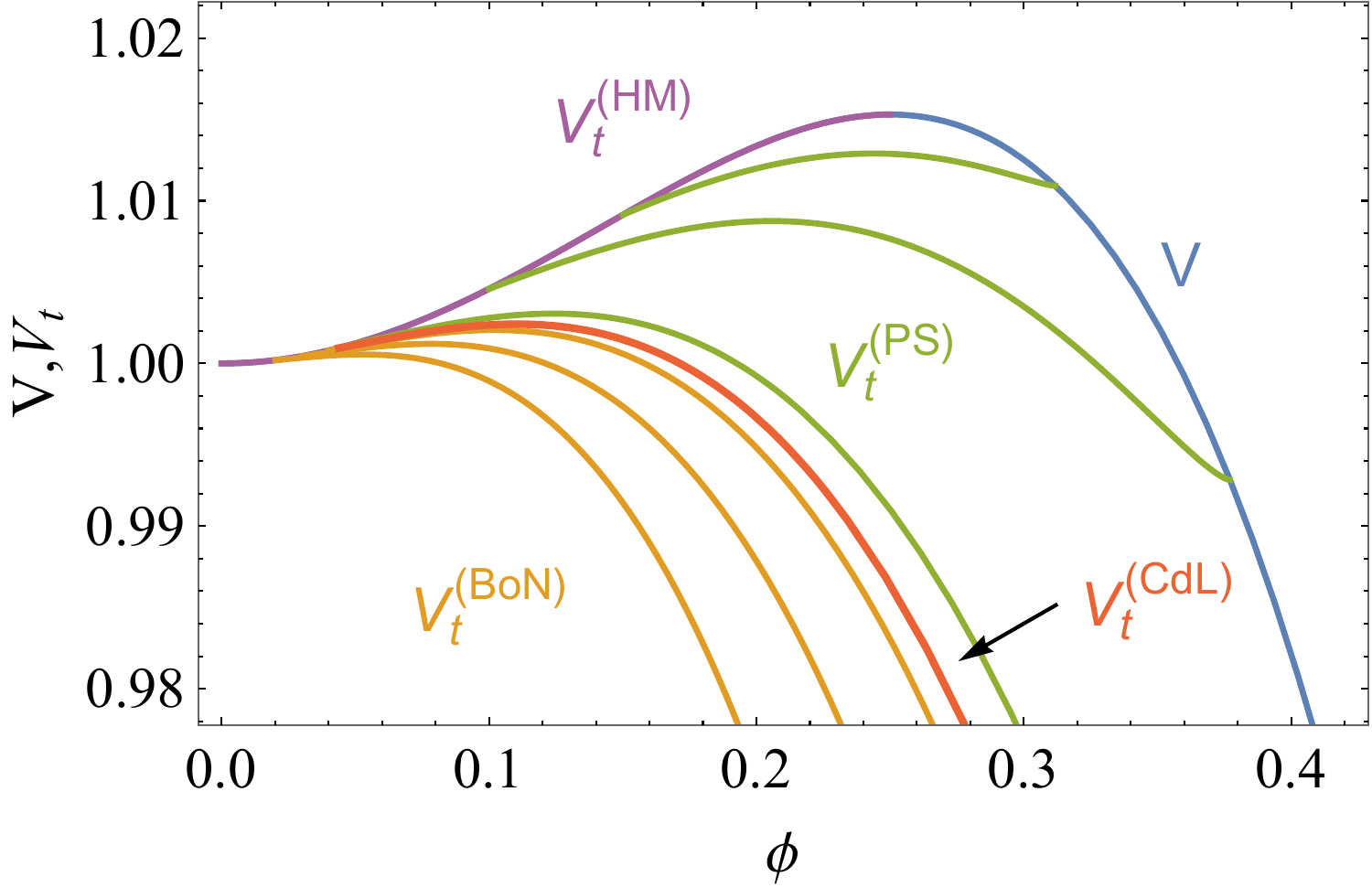}
\caption{\em Potential with a dS false vacuum and tunneling potentials $V_t(\phi_i;\phi)$ of different types: Hawking-Moss (purple); pseudo-bounces for $\phi_{0+}<\phi_i<\phi_B$ (green); CdL bounce for $\phi_i=\phi_{0+}$ (red) and BoNs for $\phi_i<\phi_{0+}$ (orange). } 
\label{fig:dSlowshoot}
\end{figure}

While in the Euclidean approach one solves the bounce equations using an overshoot/undershoot search for $\phi_0$ until the boundary conditions are satisfied, in the $V_t$ approach it is better to start the search shooting from the false vacuum (or its neighbourhood for dS vacua). Figs.~\ref{fig:Minklowshoot} and \ref{fig:dSlowshoot} show what one typically finds by doing that, for a Minkowski and dS false vacuum respectively (the AdS case is qualitatively similar to the Minkowski case), see \cite{BEHS}. 

\begin{figure}[t!]
\includegraphics[width=0.49\textwidth]{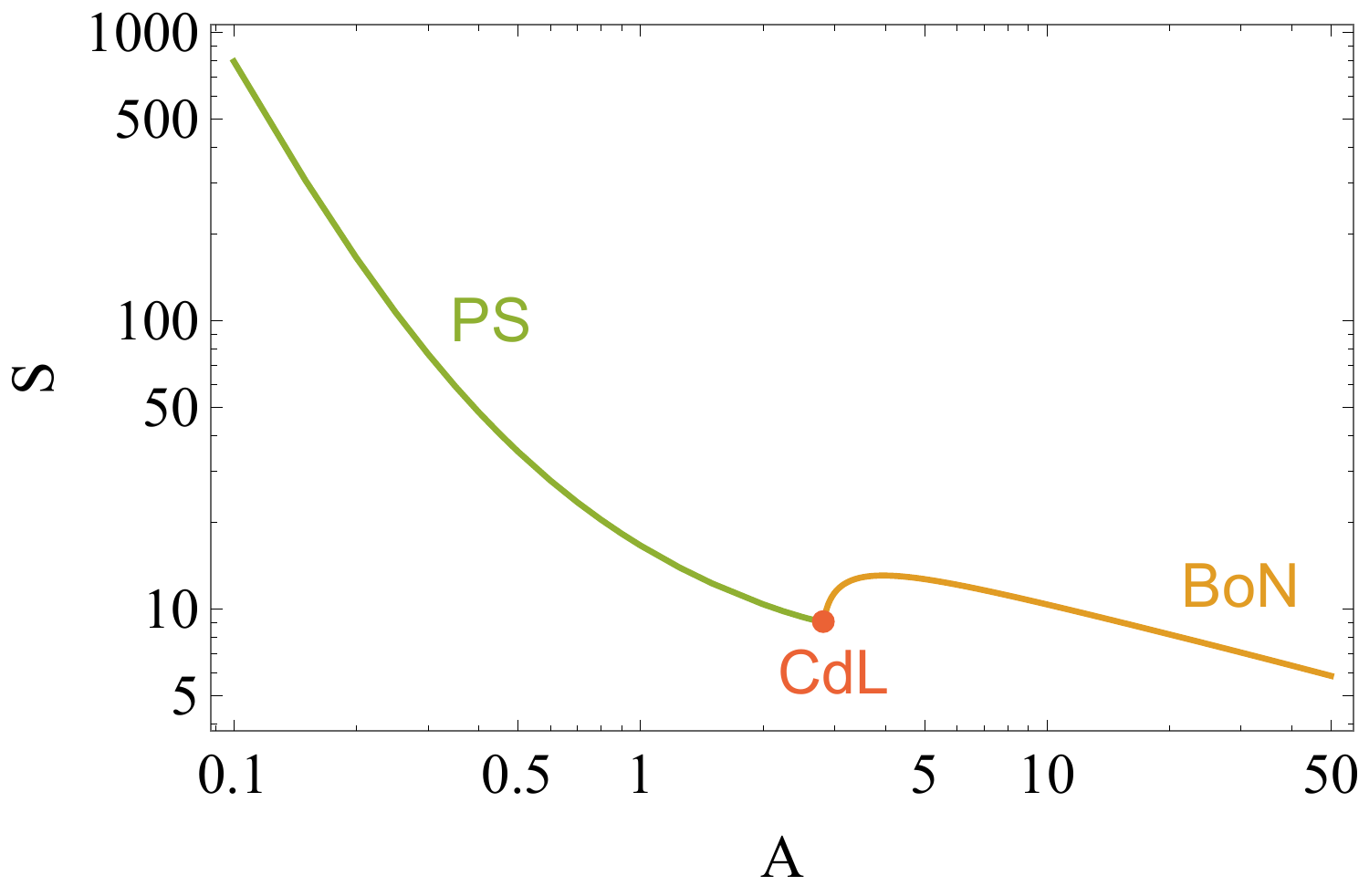}\hspace{0.3cm}
\includegraphics[width=0.46\textwidth]{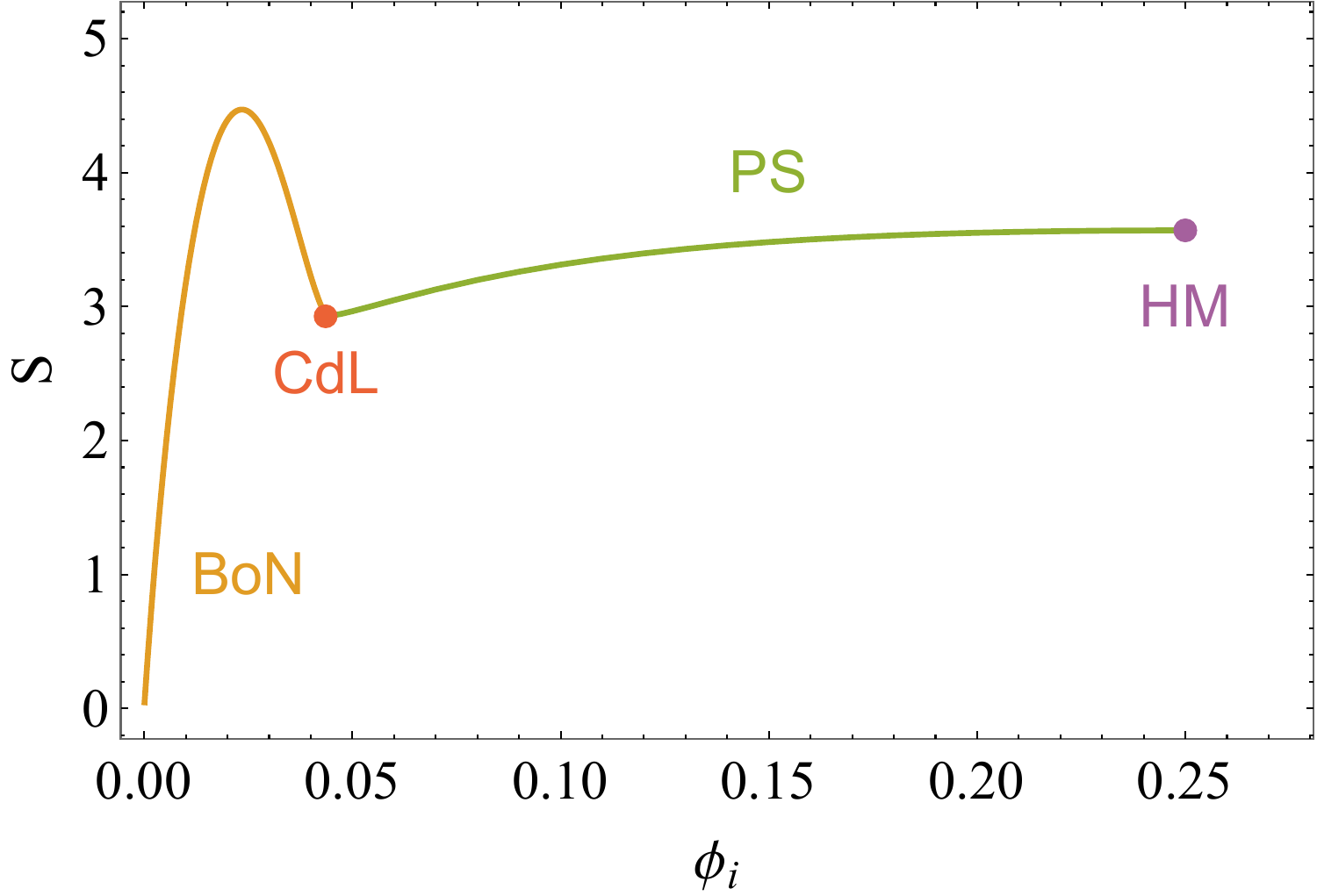}
\caption{\em Tunneling action $S$ with labels/colors indicating different types for the potentials considered in Figs.~\ref{fig:Minklowshoot} (left) and \ref{fig:dSlowshoot} (right).} 
\label{fig:actions}
\end{figure}

Let us first examine the result for the Minkowski case (Fig.~\ref{fig:Minklowshoot}). One finds that the boundary conditions at $\phi_+$, as given in (\ref{BCMinkphip}), do not fix uniquely the solution and there is a continuous family of solutions that can be parametrized by an arbitrary constant $A$, that enters in the following way. One can derive an expansion of $V_t$ around $\phi_+$ by solving the equation of motion (EoM), Eq.~(\ref{EoMVt}), perturbatively. Setting $\phi_+=0$ for simplicity, one gets \cite{BEHS}
\be
V_t(A;\phi) = -\frac{m^2\phi^2}{W}-\frac{3\kappa m^2\phi^4}{16}\left(1+\frac{5}{2W}\right)+{\cal O}\left(\phi^6,\frac{\phi^6}{W}\right)\ ,
\label{VtexpMink}
\ee
where 
\be
W\equiv W\left(A^{-1/3} e^{1/A}\phi^{-2/3}\right)\ ,
\label{eq:prodLog}
\ee
with $W(x)$ the product-log function [which satisfies $W(x)e^{W(x)}=x$ and has the large $x$
expansion $W(x)=\log x+ (1-\log x)\log(\log x)/(\log x)+...$] and $A>0$ a free parameter.\footnote{For AdS, one gets instead 
$V_t(A;\phi) = V_+ +(1/2) m_t^2\phi^2 -A \phi^{2+\alpha}+...$
with $m_t^2=(3/2) \kappa V_+[1+\sqrt{1-4m^2/(3\kappa V_+)}]$,
$\alpha=2\kappa V_+/m_t^2$ and $A$ a free parameter.} The example shown has a rather flat $\overline{V_t}$ (black dashed line) which allows for a CdL decay (red line) but one also gets further solutions (green and orange lines) which have finite action (see Fig.~\ref{fig:actions}) and can have
a physical meaning. This section focuses on the solutions above the CdL one, called pseudo-bounces and first discussed in \cite{PS} and the next section discusses the solutions lying below the CdL one, which can be interpreted as bubbles of nothing (whenever $\phi$ is a geometric modulus).

For the case of a dS vacuum, one can solve the EoM for $V_t$ using as free parameter the initical point $\phi_i$ where $V_t$ deviates from $V$, with boundary conditions as in (\ref{BCdSphi0p})
\be
V_t(\phi_i)=V(\phi_i)\ ,\quad V_t'(\phi_i)=\frac34 V'(\phi_i)\ ,
\ee
The example shown in Fig.~\ref{fig:dSlowshoot} allows for a CdL decay (red line) and also has pseudo-bounce solutions above it (with $\phi_i>\phi_{0+}$)
all the way up to the Hawking-Moss solution (with $\phi_i=\phi_T$)
as well as potential BoN solutions below the CdL $V_t$.

\begin{figure}[t!]
\begin{center}
\includegraphics[width=0.48\textwidth]{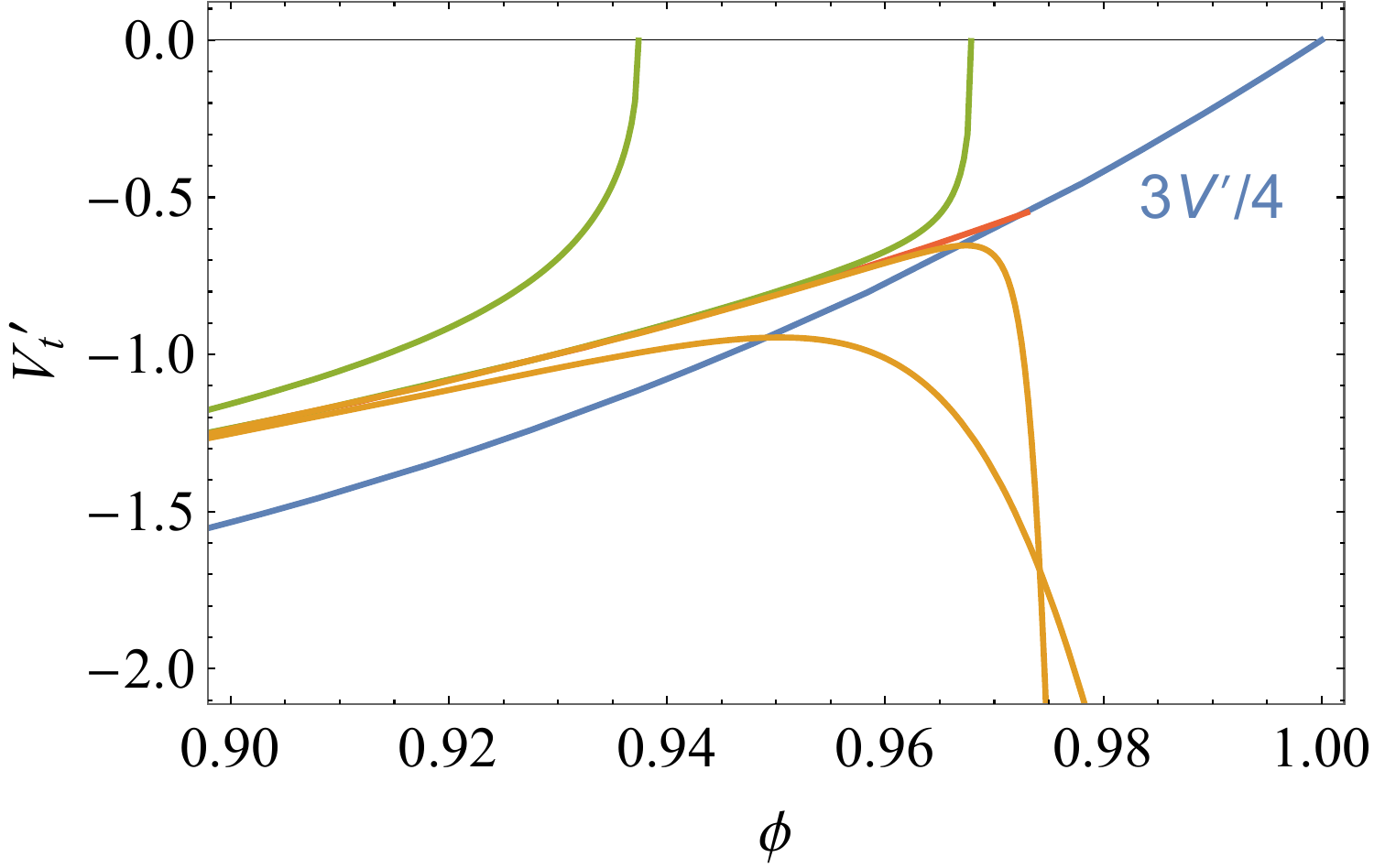}
\caption{\em Example of the behaviour of $V_t'$ near the end point $\phi_0$ for $V_t$ solutions close to the CdL one (in red), which ends with $V_t'=3V'/4$. Pseudobounce solutions (green) are driven to $V_t'=0$ while BoN solutions (orange) diverge to $V_t'=-\infty$.} 
\label{fig:Vtp}
\end{center}
\end{figure}

\begin{figure}[t!]
\includegraphics[width=0.48\textwidth]{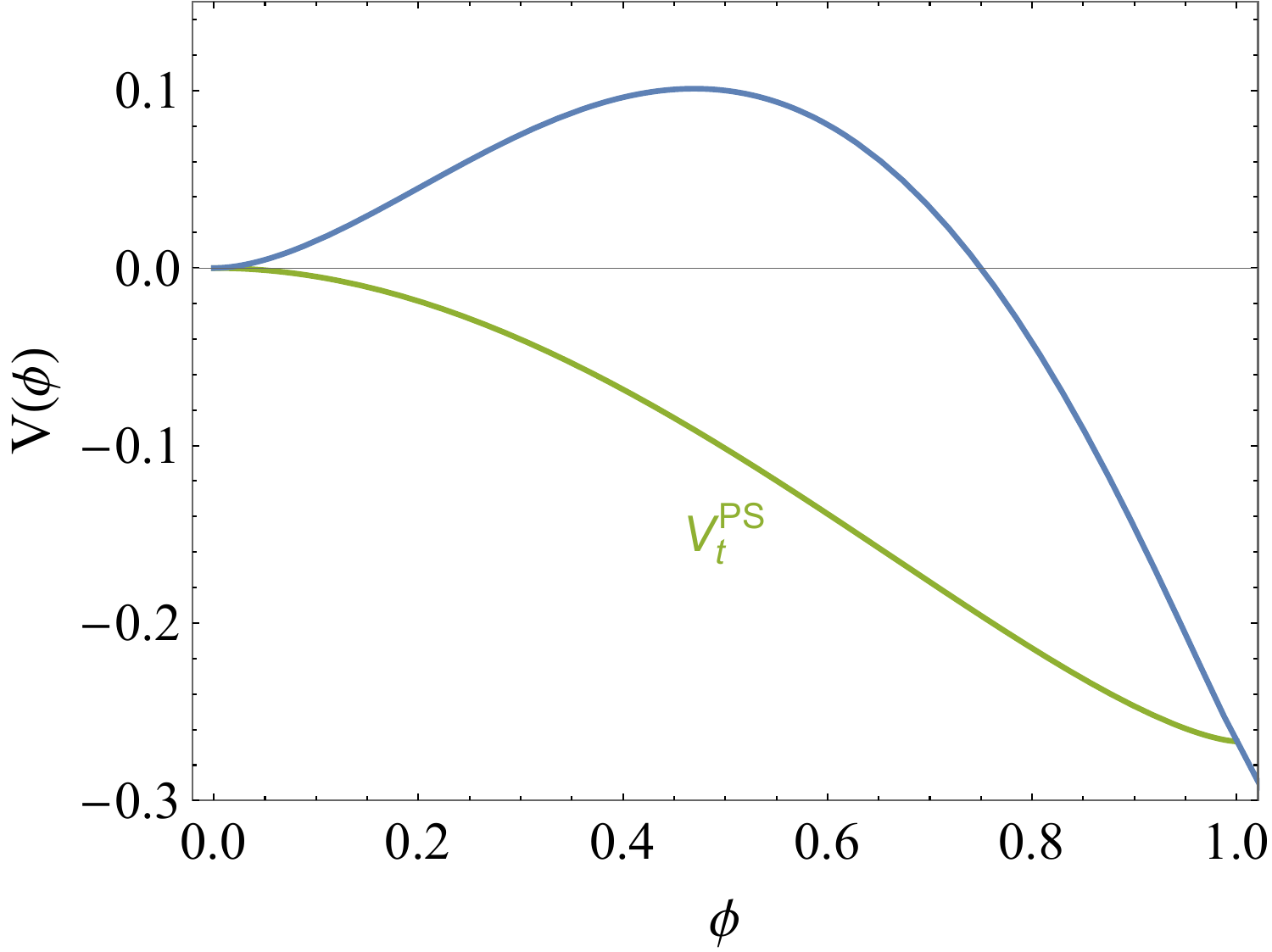}\hspace{0.3cm}
\includegraphics[width=0.46\textwidth]{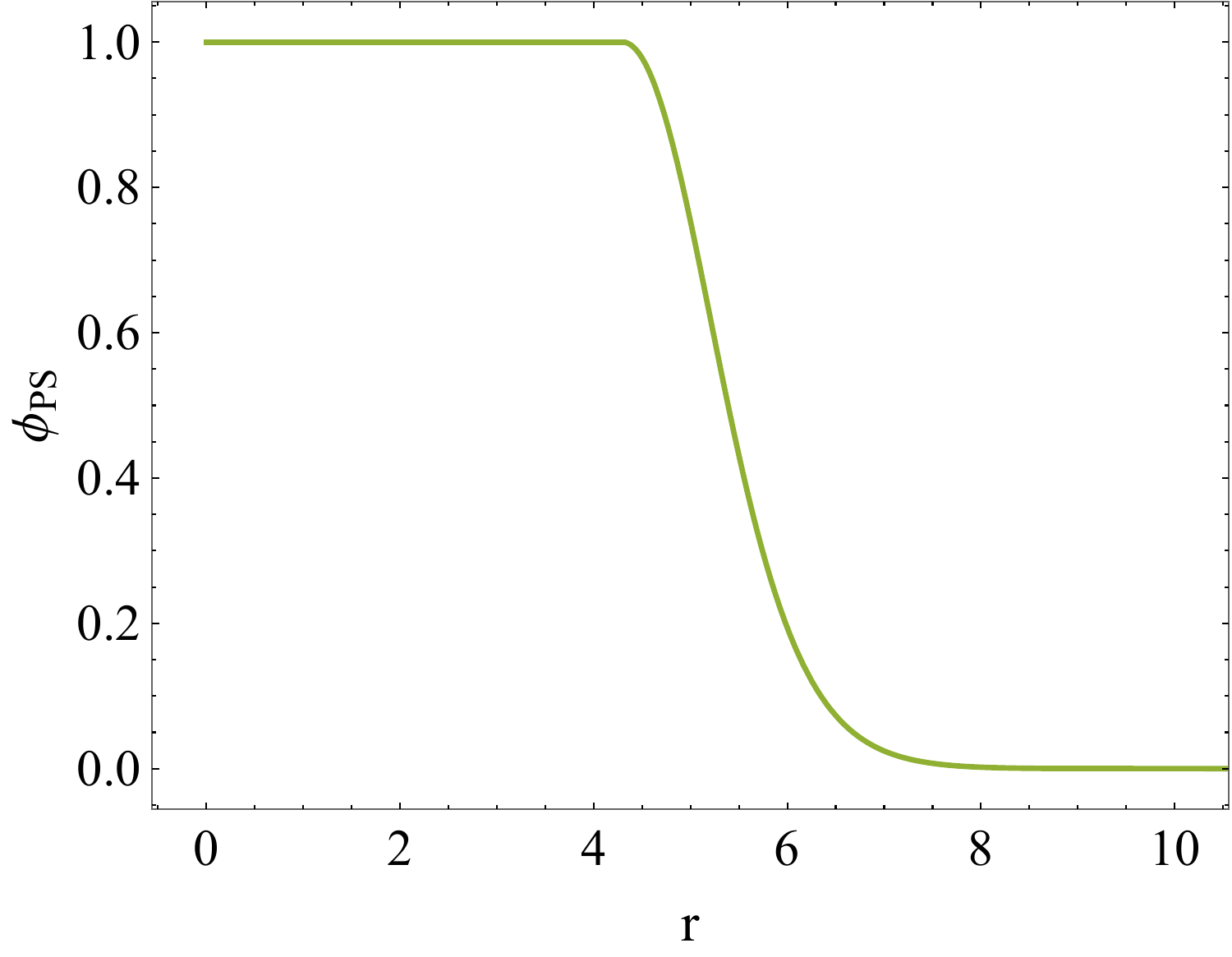}
\caption{\em A pseudo-bounce tunneling solution (left) and the corresponding Euclidean pseudo-bounce profile (right). This example correspond to an analytical case without gravity studied in \cite{PS}, with $V_t'=-\phi\sqrt{1-\phi}$.} 
\label{fig:PS}
\end{figure}

The pseudo-bounce solutions correspond to solutions of the variational problem for $V_t$ if the end point of the tunneling is held fixed. Due to this restriction, pseudo-bounces do not correspond to extremals of the action, and indeed
Fig.~\ref{fig:actions} shows that the action has non-zero slope along the pseudo-bounce family of solutions. The only stationary points are at the end of such family: the CdL solution giving the minimum value (when it exists) and the Hawking-Moss solution given the maximal value for the dS case. It can be shown \cite{PS} that the pseudo-bounces satisfy at $\phi_0$ a different boundary condition that true bounces, having $V_t'(\phi_0)=0$, see Fig.~\ref{fig:Vtp}. This can be related to the fact that the Euclidean profile of pseudo-bounces has a central core with
$\phi(r)=\phi_0$ extending out to some finite radius, see \cite{PS} for details and Fig.~\ref{fig:PS} for an example.

Pseudo-bounces are perfectly valid vacuum decay channels but, in general, as they have a tunneling action larger than the CdL solution, they can be ignored as less probable decay channels. However, there are at least two situations in which they can be relevant: 1) if the slope of the action along the pseudo-bounce family of solutions is small, they are not very much suppressed compared to CdL and one should integrate over them to get a reliable estimate of the total decay rate; 2) in some potentials the CdL solution is pushed out to infinity (in other words, the minimum of the action for the pseudo-bounce family is only reached at asymptotic infinity) and thus there is no proper bounce describing vacuum decay. In such case, one again should integrate over the family of pseudo-bounces to get the decay rate. Traditionally, faced with a situation like 2), people resorted to the use of constrained instantons \cite{CI} modifying the theory in some way to force the appearance of a bounce.  Such procedure normally results in some degree of arbitrariness and can be interpreted as an attempt at tracking the bottom of the valley in configuration space that slopes towards the asymptotic CdL bounce.
Pseudo-bounces are, by construction, exactly tracking the bottom of such valley, without any arbitrariness left [think in terms of transversal slices of the valley with constant $\phi_0=\phi_B(0)$]. Indeed, any constrained instanton with core central value $\phi_0$ would have a tunneling action higher than (or at most equal to) the action of the pseudo-bounce for that $\phi_0$ value. 

The $V_t$ approach is once again remarkable in giving in such a natural way the pseudo-bounce solutions (which look somewhat contrived in the Euclidean approach). More precisely, the pseudo-bounce $V_t$ solutions
automatically satisfy $V'_t(\phi_0)=0$ when they reach $V$ (see Fig.~\ref{fig:Vtp}), in such a way that $r$ from (\ref{r}) [or $\rho$ from (\ref{rho}) in the case with gravity] tend to a finite value which corresponds to the inner radius of the central core of the Euclidean profile (see Fig.~\ref{fig:PS}). Moreover, their tunneling action is given by the usual action $S[V_t]$ in (\ref{SVt}) without any modification [while in the Euclidean formalism the pseudo-bounce action gets contributions from the $\phi(r)$ constant core (where the EoM is not satisfied) plus the outer region where $\phi(r)$ does solve the EoM].

\section{Bubbles of Nothing\label{sec:BoN}}

Bubbles of nothing (BoNs) are a qualitatively new decay process that can occur 
in theories with compact extra dimensions and was first discussed by Witten \cite{BoN} for the $\mathbb{M}^4\times S^1$ Kaluza-Klein (KK) model. A BoN describes a tunneling from the homogeneous spacetime with constant size of the extra dimension to an spacetime with a bubble/hole with the size of the compact dimension vanishing at the surface of the bubble, and no spacetime in the interior (thus the name of BoN). 
After nucleation, the BoN expands destroying the parent spacetime.


BoNs are also important in the Swampland program,  that aims at understanding which effective field theories can be consistently coupled to quantum gravity \cite{Swamp}. The Swampland conjecture in \cite{Ooguri:2016pdq,Freivogel:2016qwc} claims that non-supersymmetric vacua are at best metastable, and BoN decays have been postulated as a universal decay channel for all non-supersymmetric compactifications \cite{Blanco-Pillado:2016xvf}. The generality of BoN decay is supported by the Swampland Cobordism Conjecture \cite{CobConj}, which claims that all consistent quantum gravity theories must be cobordant between them, that is, there must exist a domain wall separating them. In particular, these theories  must admit a cobordism to nothing and BoNs are such configuration, with spacetime ending smoothly on the BoN core. In other words, the Cobordism Conjecture implies that BoN decays should always be topologically allowed. 

Interestingly, BoNs admit an effective $4d$ description as singular Coleman-De Luccia  
Euclidean bounces of the modulus field $\phi$ that controls the size of the compactification \cite{DFG}. This bottom-up approach is quite useful to study the impact of the potential $V(\phi)$, that must be present in realistic models to stabilize $\phi$, and has been used  to get some of the conditions  $V(\phi)$ should satisfy for the existence of BoNs \cite{DGL}. Clearly, such $4d$ reduction offers the possibility of using tunneling potentials to describe BoNs. As it is shown below, BoNs can be described by $V_t$'s which are unbounded in the region where the extra dimension shrinks to zero. 
The $V_t$ technique allows to explore efficiently what different types of BoNs are possible. In this way,  Ref.~\cite{BEHS} identified four BoN types with characteristic asymptotics as $\phi\to\infty$ (the BoN core) corresponding to different $4+n$ origins (depending on the compact geometry and possible presence of a UV defect). 

To see how this works in more detail, let us examine the simple BoN discussed by Witten. The 5d KK spacetime (4d Minkowski $\times S^1$) is unstable against semiclassical decay via the nucleation of a BoN, described by the instanton metric
\be
ds^2= \frac{dr^2}{1-\frac{\mathcal{R}^2}{r^2}}+r^2d\Omega_3^2 + R_{KK}^2 \left(1-\frac{\mathcal{R}^2}{r^2}\right)d\theta^2\ ,
\label{BoNmetric}
\ee
where $R_{KK}$ is the KK radius, $\mathcal{R}$ is the size of the nucleated bubble,  $r\in (\mathcal{R},\infty)$, and $\theta\in (0,2 \pi)$ parametrises the KK circle. 
This instanton solution, analytically continued to Lorentzian signature \cite{BoN},  describes the tunneling from the homogeneous $\mathbb{M}^4\times S^1$ (reached in the $r\to\infty$ limit) to a spacetime in which the radius of the 5th dimension shrinks to zero (as $r\to \mathcal{R}$).
This BoN ``hole''  at $r=\mathcal{R}$ then expands  and destroys the KK spacetime.   
The BoN decay rate per unit volume  is $\Gamma/V\sim e^{-\Delta S_E}$, with $\Delta S_E=(\pi m_P R_{KK})^2$ being the difference between the Euclidean action of the bounce and that of the KK vacuum. 

The 5d BoN \eqref{BoNmetric} can be reduced to a 4d description \cite{DFG} by integrating the theory over the 5th dimension $\theta$, and introducing a modulus scalar $\phi$ with
\be
e^{-2\sqrt{2/3}\,\phi}\equiv 1-R^2_{KK}/r^2\ .
\ee 
 Furthermore,
a Weyl rescaling puts the BoN metric into CdL form (\ref{CdLmetric})
with 
\be
\frac{d\xi}{dr}\equiv \frac{1}{(1-R^2_{KK}/r^2)^{1/4}}\ ,\quad\quad
\rho(\xi)^2\equiv r^2(1-R^2_{KK}/r^2)^{1/2}\ .
 \ee
These two operations map the 5d instanton
into a CdL metric and a field profile, $\phi(\xi)$, with the BoN core at $\phi\to\infty$ ($\xi\to 0$) and the KK vacuum at $\phi\to 0$ ($\xi \to \infty$). 
The resulting CdL solution is not of the standard form as the field diverges at 
$\xi=0$. Nevertheless, its Euclidean action 
is finite and equal to Witten's (after including a boundary term of 5d origin). Explicitly, one can show that the tunneling exponent $\Delta S_E$ can be written as
\be
\Delta S_E=-2\pi^2\int_0^{\infty} \rho^3 V d\xi -\left.\pi^2\sqrt{\frac{2}{3\kappa}}\rho^3\dot\phi\,\right|_{\xi=0}\ .
\label{FinalSE}
\ee 
This result takes the standard CdL form (for decays out of a Minkowksi false vacuum) plus an additional term evaluated at $\xi=0$, which is a purely $5d$ input.

Finding the BoN description  
in the $V_t$ approach is straightforward,  using $V_t=V-\dot\phi^2/2$. One gets
\be
V_t(\phi) = -
(6/R^2_{KK})
\sinh^3(\sqrt{2/3}\,\phi)\ ,
\label{WBoN}
\ee 
with $V_t(0)=0$ and $V_t(\phi\rightarrow\infty)\sim - e^{\sqrt{6}\phi}$ ,
so that $V_t$ diverges to $-\infty$ at $\phi\rightarrow \infty$. This is a generic property of the $V_t$'s of BoNs.

Furthermore, the action in the $V_t$ formalism, Eq. (\ref{SVtg}), gives the correct result without the need of additional boundary terms.  Explicitly,
using
\be
D(\phi)=\frac{6}{R_{KK}^2}\sqrt{\frac{6}{\kappa}}\sinh^2(\sqrt{2\kappa/3}\phi)\ ,
\ee
the action density $s(\phi)$ is simply
\be
s(\phi)=\frac{\pi^2R_{KK}^2}{2}\sqrt{\frac{3}{2\kappa}}\mathrm{sech}^4(\sqrt{\kappa/6}\,\phi)\ ,
\ee
which is finite everywhere and integrates to the correct result:
\be
S[V_t]=\int_0^\infty s(\phi)d\phi = (\pi m_P R_{KK})^2\ ,
\label{SWBoN}
\ee
without having to include boundary terms as in the Euclidean approach.
The agreement between the simple action $S[V_t]$ given by (\ref{SVtg}) and the Euclidean action result holds in general, not only for Witten's BoN (for the proof, see Appendix~B of \cite{BEHS}).

So, once again one finds a case in which the tunneling potential formalism behaves better than one would have expected. In the same way that the Euclidean action calculation requires the inclusion of boundary terms that come from the extra compact dimensions of the theory, it would have been only natural to expect that also the $V_t$ action calculation would receive some boundary corrections of similar origin. However, no such corrections are necessary once the problem is formulated in field space rather than in spacetime, which is certainly an attractive feature.

\begin{figure}[t!]
\begin{center}
\includegraphics[width=0.48\textwidth]{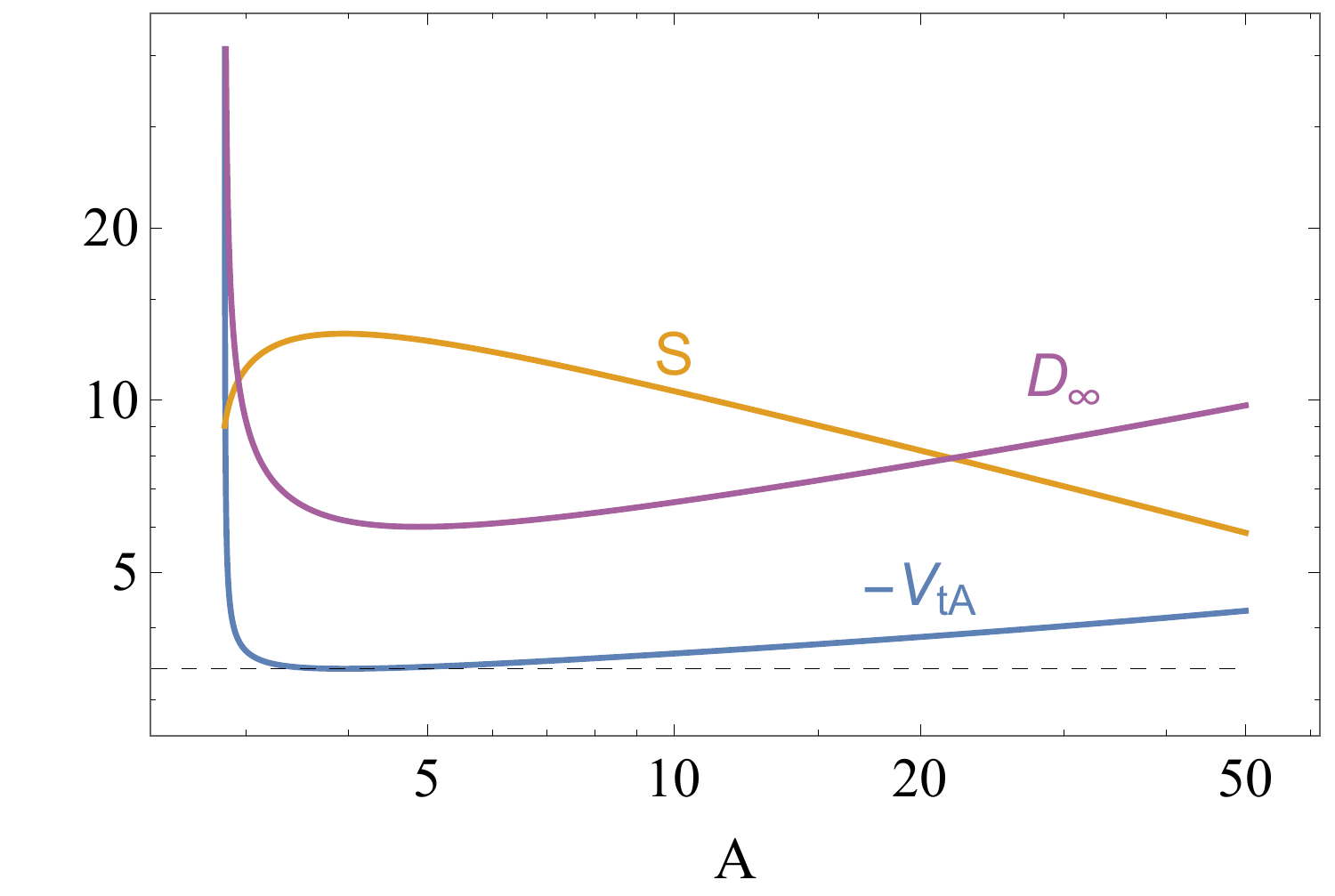}
\caption{\em For the potential of Fig.~\ref{fig:Minklowshoot} and the BoN range of the free parameter $A$, action $S$ and prefactors $V_{tA}, D_{\infty}$ controlling the asymptotic $\phi\to\infty$ behaviors $V_t\sim V_{tA}e^{\sqrt{6\kappa}\phi}$ and  $D\sim D_\infty e^{\sqrt{8\kappa/3}\phi}$. The black dashed line gives the lower bound on $-V_{tA}$.} 
\label{fig:VtADA}
\end{center}
\end{figure}

The $V_t$ formalism is thus well suited to analyze in a very direct way  the general properties of BoNs in the presence of a potential for the modulus field, without the need of descending to the level of the field and metric profiles. One finds that the $\phi\to\infty$ asymptotics of $V_t$ and its corresponding $D$ depend on a free parameter that cannot be determined by the 4d BoN solution (thus one finds a continuous family of BoN solutions).  Once a particular $d=4+n$ theory is chosen, that constant is fixed in terms of geometric parameters of the compact space [as in Witten's BoN, see (\ref{WBoN})]. As an example, for a 5d KK ($4d$ Minkowski $\times S^1$) theory, one gets the 4d asymptotics
\be
V_t(\phi\to\infty)\simeq V_{tA}(A)e^{\sqrt{6\kappa}\phi}\ ,\quad\quad
D(\phi\to\infty)\simeq D_{\infty}(A)e^{\sqrt{8\kappa/3}\phi}
\ee
with $V_{tA}(A)$ and $D_{\infty}(A)$ some functions of a free parameter $A$, an example of which is shown in Fig.~\ref{fig:VtADA}. The 5d analysis is able to explain the values of $V_{tA}$ and $D_{\infty}$ in terms of 5d fundamental quantities. In this example one gets
\be
V_{tA}=-\frac{C_1}{\kappa R_{KK}^2}\ ,\quad\quad 
D_{\infty}=\frac{C_2}{\sqrt{\kappa}{\cal R} R_{KK}}\ ,
\label{5dVtD}
\ee
where $R_{KK}$ is the Kaluza-Klein radius, ${\cal R}$ is the BoN radius and $C_{1,2}$ are numerical constants. 
Thus, from Fig.~\ref{fig:VtADA} one concludes the following: for a fixed theory (fixed $R_{KK}$), provided $-V_{tA}$ from (\ref{5dVtD}) is above the minimal value of $-V_{tA}$ shown in the figure, two possible $V_t$ solutions exist. For each one, the $D_\infty(A)$ curve fixes the corresponding BoN radius via (\ref{5dVtD}) and the $S$ curve gives the corresponding tunneling action. Whichever solution has lower $S$ will be the preferred BoN decay channel. One also sees that, if $-V_{tA}$ from (\ref{5dVtD}) lies below the minimal value of $-V_{tA}$ shown in the figure, no $V_t$ solution would exist and thus no BoN decay would be open. This dynamical quenching of BoN decays would be an interesting effect, although the region of parameter space where it is possible seems to be on the verge of validity of the 4d effective theory. Elucidating the robustness of this quenching effect requires a more detailed scrutiny. For additional results of the $V_t$ analysis of BoN decays, see \cite{BEHS}.

\section{Stability of AdS Maxima}

Another application of the tunneling potential approach is to the analysis of the stability of AdS maxima, that is, field backgrounds with negative squared mass above the Breitenlohner-Freedman bound. To be concrete, consider potentials whose low-field expansion around an AdS maximum ($V_+<0$) at $\phi_+=0$ reads
\be
V(\phi)=V_+ +\frac12 m^2 \phi^2+...\ ,
\label{V}
\ee
with $m^2<0$ but not too negative, so that $m^2>m_{BF}^2$, where $m_{BF}^2$ is the Breitenlohner-Freedman \cite{BF} (squared) mass
\be
m_{BF}^2\equiv \frac{3\kappa}{4} V_+<0\ .
\ee
Provided this Breitenlohner-Freedman (BF) bound is satisfied, gravitational curvature effects make the scalar field non-tachyonic, and thus, the homogeneous $\phi_+=0$ background is stable. Nevertheless, quantum tunneling via CdL instantons can still make such field backgrounds unstable. Before examining this question, let us first remark that the study of such problem is not just of academic interest: such negative mass states over the BF bound often appear in supergravity theories derived from strings. Moreover, the mass range
\be
m_{BF}^2<m^2<m_{BF}^2+\frac{1}{L_{AdS}^2}\ ,
\label{m2range}
\ee
where $L_{AdS}=\sqrt{3/(2\kappa |V_+|)}$ is the AdS length at the $V_+$ maximum of $V$ (for a spacetime with $d=4$ dimensions) is particularly relevant for the AdS/CFT correspondence \cite{Witten,HH}, for black hole solutions with scalar hair \cite{BHhair}, for the so-called designer gravity \cite{designer}, etc.

\begin{figure}[t!]
\begin{center}
\includegraphics[width=0.48\textwidth]{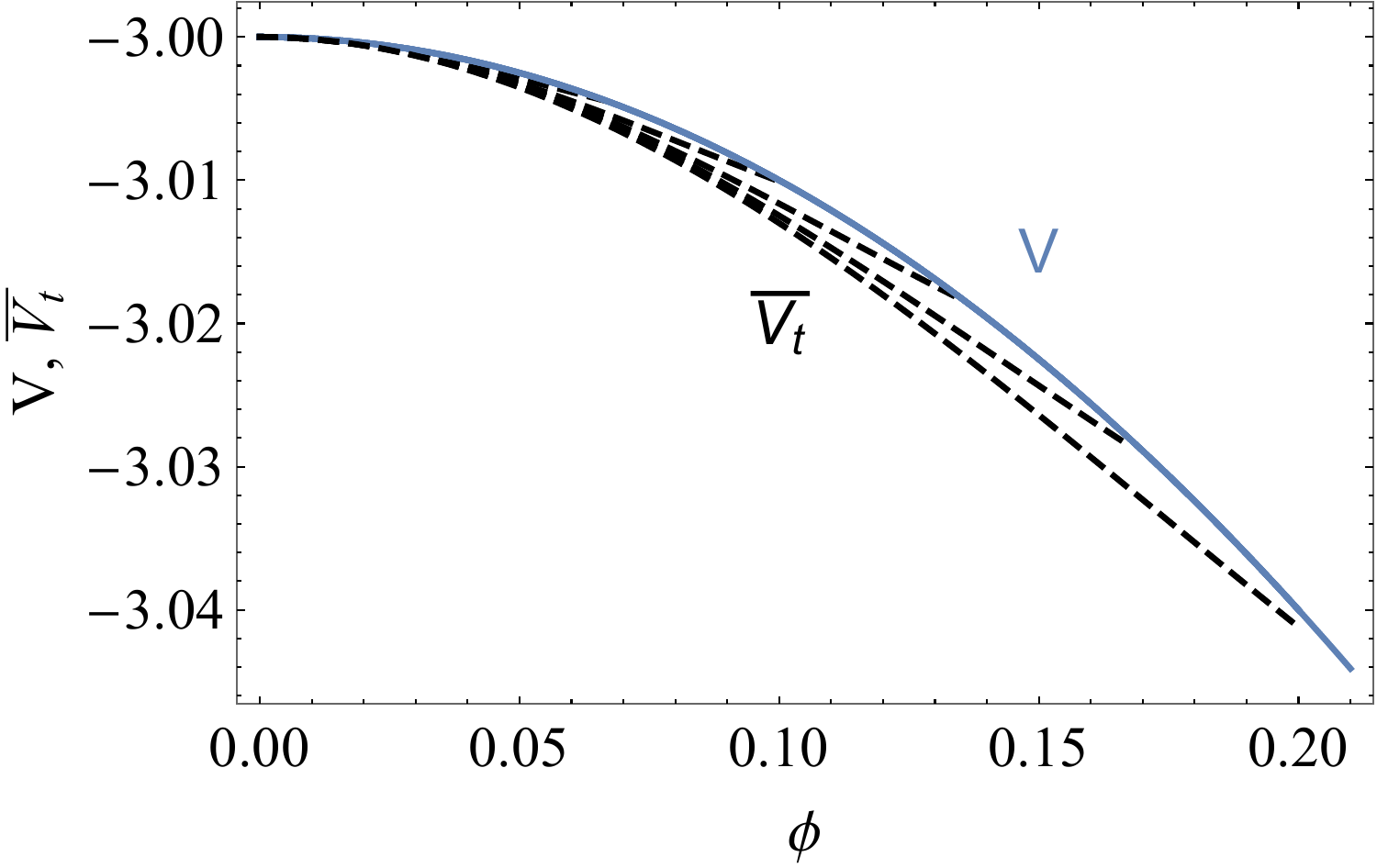}
\includegraphics[width=0.48\textwidth]{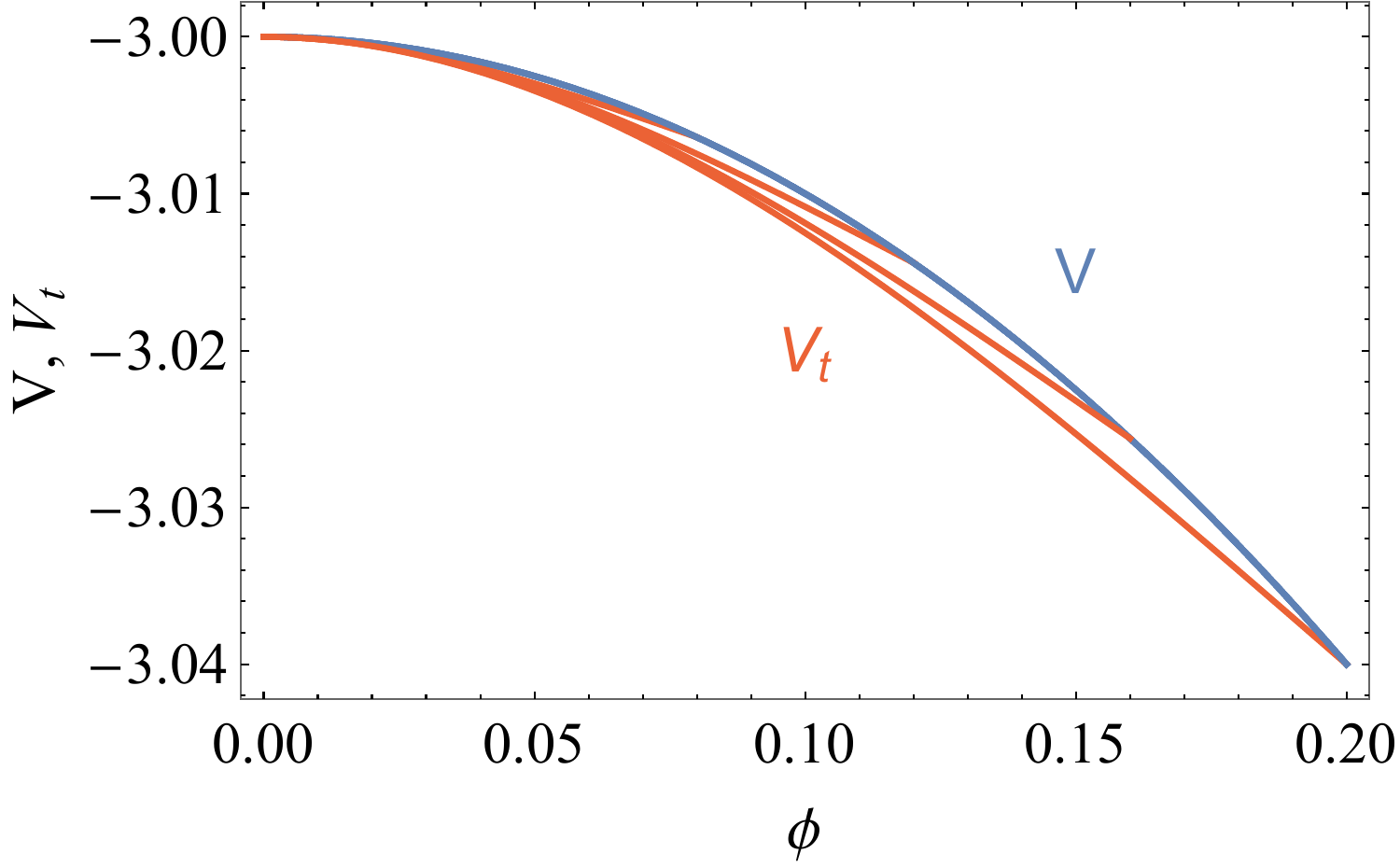}
\caption{\em For a potential with a maximum at $\phi=0$ with $V_+=-3$, Left plot: family of $\overline{V_t}$ solutions (dashed lines);
Right plot: family of CdL $V_t$ solutions (red lines).} 
\label{fig:VtcAdsMax}
\end{center}
\end{figure}

Can one use the same $\overline{V_t}$ function introduced at the end of Subsection~\ref{Vtgrav} to determine if a given AdS potential maxima can suffer CdL decay? If one tries this, solving $D^2=(\overline{V_t}')^2+6\kappa(V-\overline{V_t})\overline{V_t}=0$ with $\overline{V_t}(\phi_+)=V_+$, $\overline{V_t}'(\phi_+)=0$ and choosing the solution with $\overline{V_t}'\leq0$, one encounters a surprise: instead of a single $\overline{V_t}$ solution there is a continuous family of solutions. Indeed, suppose a given $\overline{V_t}$ solution exists that reaches $V$ at some field value $\phi_e$. One can find further solutions, starting from
some $\phi_e'$ in the interval $(\phi_+,\phi_e)$ and solving $D=0$ starting from $\phi_e'$
with $\overline{V_t}(\phi_e')=V(\phi_e')$ towards $\phi_+$. One is guaranteed to reach 
$\phi_+$ as the new solution cannot intersect neither the previous $\overline{V_t}$ solution nor the potential $V$ before reaching 
$V_+$ at $\phi_+$. Thus, all solutions of $D=0$ for any $\phi_e'$ in the interval $(\phi_+,\phi_e)$ are funnelled towards $V_+$ at $\phi_+$, which is an accumulation point. See Fig.~\ref{fig:VtcAdsMax} (left plot) for an example of this.

One can see how this family of solutions appears in the following way. Taking $\phi_+=0$ and solving $D=0$ perturbatively at small field value,
one finds \cite{EJ} that the critical tunneling potential $\overline{V_t}$ has two possible expansions. First, there is a family of $\overline{V_t}$ solutions with the expansion
\bea
\overline{V_t}_-(\phi)&=&V_+ +\frac12 m_{t-}^2\ \phi^2+B_{4-}\ \phi^4+B_{6-}\ \phi^6+...\nonumber\\
&+&\overline{A}\ \phi^{2+\overline{\gamma}_-}+\overline{A}_{2}\ \phi^{2+2\overline{\gamma}_-}+\overline{A}_{4}\ \phi^{4+\overline{\gamma}_-}+...\ ,
\label{Vtcminus}
\eea
that can be labelled by the continuous free parameter $\overline{A}$. At the lower boundary of the previous family one finds another $\overline{V_t}$ with the form
\bea
\overline{V_t}_+(\phi)&=&V_+ +\frac12 m_{t+}^2\ \phi^2+B_{4+}\phi^4+\ B_{6+}\phi^6+...\ ,
\label{Vtcplus}
\eea
and no free parameter. In the expressions above
\be
m_{t\pm}^2= \frac43 m_{BF}^2\Delta_\pm\ ,\quad\quad
2+\overline{\gamma}_\pm=\frac{3}{\Delta_\pm}\ ,
\label{mtpm}
\ee
with
\be
\Delta_\pm\equiv \frac32 \left(1\pm \sqrt{1-m^2/m_{BF}^2}\right)\ ,
\ee
and, for simplicity of notation, I use sometimes $\Delta\equiv \Delta_-$ below.
Note that, for $\overline{V_t}(\phi)$ to be real, the Breitenlohner-Freedman bound ($m^2\geq m_{BF}^2$) must be satisfied.

The coefficients of the higher order terms in (\ref{Vtcminus}) for $\overline{V_t}_+$ depend on the possible higher order terms in the potential. Adding just a quartic coupling $\lambda\phi^4/4$ to $V$ in (\ref{V}) one gets 
\be
B_{4\pm}=\frac{3\lambda-4\Delta_\pm\lambda_{\pm}}{4(3-4\Delta_\pm)}\ ,\quad\quad
B_{6\pm}=\frac{(\lambda-\lambda_{\pm})\left[\lambda+4(\Delta_\pm-1)\lambda_{\pm}\right]}{2(3-4\Delta_\pm)^2(2\Delta_\pm-1)}\ ,\ ...
\label{Bs}
\ee
where
\be
\lambda_\pm\equiv \frac{V_+\kappa^2}{4}\Delta_\pm^2=-\frac{3}{4}\Delta_\pm^2\ ,
\ee
and
\be
\overline{A}_{2}=\frac{3\overline{A}^2}{2(3-2\Delta)\Delta^2}\ , \quad \quad
\overline{A}_{4}=\frac{\overline{A}\left[9\lambda+2(4\Delta^2+3\Delta-9)\lambda_-\right]}{8(4\Delta-3)\lambda_-} \ ,\ ...
\ee

What should one conclude about the stability of such AdS maxima from the existence of not just one critical $\overline{V_t}$ but of a whole family of $\overline{V_t}$'s? Moreover, a similar analysis for $V_t$ CdL solutions shows a similar pattern, with a whole family of $V_t$ solutions. Does this mean that AdS maxima are unstable via quantum tunneling and with an infinite number of possible decay channels? The answer is negative and to understand it one should remember that the dynamics of scalar fields in AdS is crucially affected by boundary conditions at the conformal boundary. Writing the (4d) AdS metric as
\be
ds^2=-\left(1+\rho^2/L_{AdS}^2\right)dt^2+\frac{d\rho^2}{1+\rho^2/L_{AdS}^2}+\rho^2d\Omega_2^2\ ,
\ee
the asymptotic fall-off of a free scalar field of squared mass $m^2$ at $\rho\to\infty$ is obtained from the Klein-Gordon equation in AdS as
\be
\phi(\rho)\simeq \frac{\alpha}{\rho^{\Delta_-}}+\frac{\beta}{\rho^{\Delta_+}}+...
\label{phiasympt}
\ee
For $m^2>0$, $\Delta_-<0$ and one is forced to have $\alpha=0$ to avoid a divergent behaviour. However, for $m^2$ in the range (\ref{m2range}) both terms in (\ref{phiasympt}) lead to normalizable solutions. For $\alpha\neq 0$, the field has a slower fall-off at $\rho\to\infty$ which looks problematic in principle as it can  modify the asymptotic symmetries and charges of the theory and cause divergences in the energy and action of the scalar field plus gravity \cite{HMTZ}. Remarkably, charges, energy and action can be modified by appropriate scalar boundary terms to get conserved charges, and finite energy and action \cite{HMTZ} (an explicit example of this is discussed below). For this to work, it is required that $\alpha$ and $\beta$
are related by a functional relation
\be
\beta(\alpha)=W'(\alpha)\ ,
\ee
and different choices of the function $W(\alpha)$ correspond to different theories. By appropriately choosing $W(\alpha)$ one can thus tailor at will different properties of the scalar plus gravity theory (this was dubbed designer gravity \cite{designer}). 

The asymptotic behaviour of the fields as $\rho\to\infty$ is thus different in different theories and this translates into different theories corresponding to different asymptotic behaviours of $\overline{V_t}$ and $V_t$.\footnote{The asymptotics of $\overline{V_t}$ and $V_t$ can be derived \cite{EJ} using the dictionary between Euclidean and $V_t$ formalisms.} The resolution of the stability puzzle for AdS maxima is thus that different theories (thus different boundary conditions of the scalar field at $\rho\to\infty$) select particular members of the families of $\overline{V_t}$ and $V_t$ which are the relevant ones for that theory.\footnote{This is similar to what happens for BoNs, where the particular $4+n$ theory selects particular solutions among the continuous family of BoNs, see Sect.~\ref{sec:BoN}.} 

To see all of this at work, let us look at a simple example. Set $V_+=-3$, $\kappa=1$
(so that $L_{AdS}=1$), and consider a scalar field with $m^2$ in the range (\ref{m2range}) taking in particular $\Delta=1$ [so that (\ref{phiasympt}) is simply $\phi(\rho)\simeq \alpha/\rho+\beta/\rho^2+...$].\footnote{For more general results in the range $1/2<\Delta<3/2$, see \cite{EJ}.} Restricting the analysis to small field values (close to the AdS maximum at $\phi_+=0$) one can solve analytically for the CdL solutions, both in Euclidean and $V_t$ formulations. Both analyses are presented below in order to contrast them and to highlight another unexpected feature of the $V_t$ approach, .

\subsection{CdL Instantons. Euclidean Approach}
Small-field Euclidean CdL instantons with $\phi(0)=\delta\ll 1$ and $\delta$ free can be found as 
\be
\phi(\xi)=\delta\ {\rm sech}^2(\xi/2)+{\cal O}(\delta^2)\ .
\label{phixi}
\ee
with the metric function $\rho(\xi)=\sinh\xi+{\cal O}(\delta^2)$\ .
It is convenient to use $\rho$ as independent variable instead of $\xi$,
to get the instanton profile as
\be
\phi(\rho) = 2\delta\frac{\sqrt{1+\rho^2}-1}{\rho^2}+{\cal O}(\delta^2)\ .
\ee
The large-$\rho$ expansion of the previous result  gives
\be
\phi(\rho)\simeq \frac{2\delta}{\rho}-\frac{2\delta}{\rho^2}+{\cal O}(1/\rho^3)\ ,
\ee
which conforms to the expected $\Delta=1$ behaviour, with 
\be
\alpha=2\delta\ , \quad \beta=-2\delta\ .
\ee
This determines a trajectory in the $\{\alpha,\beta\}$ plane parametrized by $\delta$. If that curve intersects with the "theory curve" $\{\alpha,W'(\alpha)\}$, the intersection point corresponds to a CdL decay channel of the AdS maximum in the theory defined by $W(\alpha)$ \cite{designer}. If there is no intersection, then the AdS maximum is stable and one can derive a positive energy theorem for that scalar background \cite{designer}.

It is also instructive to calculate the Euclidean action for such CdL tunnelings.
As usual, the Euclidean tunneling action is the difference between the instanton action and the AdS background action for the false vacuum. The instanton action is the sum of three contributions \cite{HH}: the usual gravitational action
\be
S_G=\int d^4x \sqrt{g}\left[-\frac{R}{2\kappa}+\frac12 (\nabla\phi)^2+V(\phi)\right]\ ,
\ee
which can be rewritten, for the $O(4)$ symmetric instanton,  as
\be
S_G=-2\pi^2\int_0^\infty\rho^3V(\phi)d\xi\ ;
\ee
the boundary Gibbons-Hawking-York term, which can be written as
\be
S_{GHY}=-\frac{6\pi^2}{\kappa}\int_0^\infty d[\rho^2\dot\rho]\ ;
\ee
and, as mentioned above, an additional surface term, necessary  due to the scalar field falling off at infinity slower than usual,
\be
S_{S}= \oint [(\nabla\phi)^2-m^2\phi^2]\ ,
\ee
which can be rewritten as
\be
S_{S}=\frac{\pi^2}{3}\int_0^\infty d[\rho^3(\dot\phi^2-m^2\phi^2)]\ .
\ee
The AdS background action is just $S_G+S_{GHY}$ evaluated at the false vacuum $\phi_+=0$ with $\rho=\rho_+=\sinh(\xi L_{AdS})/L_{AdS}$. The total Euclidean tunneling action is then
\be
\Delta S_E\equiv S_G+S_{GHY}+S_S-S_{AdS}\ ,
\label{DSE}
\ee

Let us calculate all of the above terms for our small field solutions above.
To use $\rho$ as integration variable one simply needs to use
$d\xi=d\rho/\dot\rho$, with $\dot\rho\equiv d\rho/d\xi$ given by (see footnote~\ref{foot})
\be
\dot\rho= \sqrt{\frac{1-\rho^2V(\phi(\rho))/3}{1-\rho^2(d\phi/d\rho)^2/6}}\ .
\label{xitorho}
\ee 
In this case, one gets
\be
\dot\rho=\sqrt{1+\rho^2}+\frac{\left(\sqrt{1+\rho^2}-1\right)^2}{3\rho^4\sqrt{1+\rho^2}}\left[\left(\sqrt{1+\rho^2}-1\right)^2+2\rho^2\right]\delta^2 +{\cal O}(\delta^3)\ .
\ee
The explicit results for the different contributions to the tunneling action (\ref{DSE}) are given below, using a UV cutoff $\rho_\infty$ in the $\rho$ integrals to see explicitly the cancellation of divergences. For the instanton contributions one gets
\bea
S_G&=&2\pi^2\left[2+(\rho_\infty^2-2)\sqrt{\rho_\infty^2+1}+\left(-2+\frac{\rho_\infty^2+2}{\sqrt{\rho_\infty^2+1}}\right)\delta^2\right]+{\cal O}(\delta^3)\ ,\nonumber\\
S_{GHY}&=&-2\pi^2\left\{3\rho_\infty^2\sqrt{\rho_\infty^2+1}+\frac{\left(1-\sqrt{\rho_\infty^2+1}\right)^2}{\rho_\infty^2\sqrt{\rho_\infty^2+1}}\left[2\rho_\infty^2+\left(1-\sqrt{\rho_\infty^2+1}\right)^2\right]\delta^2
\right\}+{\cal O}(\delta^3)\ ,\nonumber\\
S_S&=&\frac{4\pi^2}{3\rho_\infty^2}
\left(1-\sqrt{\rho_\infty^2+1}\right)^2\left[2\rho_\infty^2+\left(1-\sqrt{\rho_\infty^2+1}\right)^2\right]\delta^2
+{\cal O}(\delta^3)\ ,
\eea
while for the AdS background one has
\be
S_{AdS}=S_{G,AdS}+S_{GHY,AdS}=4\pi^2\left[1-(\rho_\infty^2+1)^{3/2}\right]\ .
\ee
Putting all the pieces together, the tunneling action is
\bea
\Delta S_E &=&\lim_{\rho_\infty\to\infty}\frac{4\pi^2\left(\sqrt{\rho_\infty^2+1}-1\right)^2}{3\rho_\infty^3\sqrt{\rho_\infty^2+1}}\left[3\rho_\infty(\rho_\infty-1)\left(\sqrt{\rho_\infty^2+1}-1\right)\right.\nonumber\\
&&\left.+2\rho_\infty^2-(3\rho_\infty+1)\left(\sqrt{\rho_\infty^2+1}-1\right)^2
\right]\delta^2+{\cal O}(\delta^3)\nonumber\\
&=&\frac{4\pi^2}{3}\delta^2+{\cal O}(\delta^3)\ .
\eea
Remarkably, $\Delta S_E$
is finite, with all the infinities appearing in individual contributions cancelling out (this cancellation happens in general, see e.g. Appendix of \cite{HH}).

\subsection{CdL Instantons. $\bma{V_t}$ Approach}
In the tunneling potential approach, the small-field solution (with $V=-3-\phi^2$) turns out to be simply
\be
V_t(\phi) \simeq -3 -\frac32 \phi^2 +\frac{1}{2\delta}\phi^3\ .
\label{Vtdelta}
\ee
The expression for the tunneling action (\ref{SVtg})
does not need to be supplemented by additional terms like the Euclidean one. Plugging (\ref{Vtdelta}) in it, and using $\phi_0=\delta$ (as this is the field value at which $V_t=V$), one immediately gets
\be
S=\frac{4\pi^2}{3}\delta^2+{\cal O}(\delta^3)\ ,
\ee
in agreement with the Euclidean result. This agreement is general and, once again, it is an unexpected result of the $V_t$ approach, in the same way that no additional boundary terms were needed to calculate the BoN tunneling actions (see Sect.~\ref{sec:BoN}). 

In the $V_t$ approach, the asymptotic behaviour of  $V_t(\phi\to\phi_+)$ can be directly related to the asymptotic behavior of the scalar $\phi$ as $\rho\to\infty$.
The constants $\alpha,\beta$ are related to the free parameters that label the $\overline{V_t}$ and $V_t$ solutions \cite{EJ}.
Different theories give different asymptotics and this is what selects the relevant $V_t$ among the full family of possible solutions. If the selected $\overline{V_t}$
intersects the potential away from the maximum, then (as in the case of the stability of vacua with $m^2>0$) the maximum is unstable against CdL decay. Otherwise the maximmum is stable and one can derive a positive energy theorem \cite{EJ}.

\section{Conclusions}

As I have tried to illustrate, the tunneling potential formalism, used to calculate the actions suppressing false vacuum decay in QFT, gives us more than we had reason to expect. Apart from having a complementary approach to an important problem, which is always good to have, one finds a number of instances in which the $V_t$
approach is extremely kind to us. 

To summarize such cases: the first unexpected and nice property is that the $V_t$ that extremizes the action functional $S[V_t]$ corresponds to a minimum rather than a saddle point. Beyond its practical use for numerical calculations, this property is of theoretical interest by its own right.
The second unexpected property is that the very same $S[V_t]$ applies in exactly the same form to any kind of false vacua, be it AdS, Minkowski or dS. This happens in spite of the fact that the dS case, being the $V_t$ solution  qualitatively different, involves an extension of the $V_t$ function to a range in which the link $V_t=V-\dot\phi^2/2$ is not applicable (in particular this is the complete range of the solution when only the Hawking-Moss instanton exists). One would have expected that $S[V_t]$ would need some patching up to cover properly dS false vacua (and Hawking-Moss transitions in particular). But, no, the simple $S[V_t]$ valid for AdS or Minkowski does the job nicely. A third instance is the ease with which solutions with physical significance follow from the formalism. That is precisely what happens with pseudo-bounces and bubbles of nothing, which jump at you from simply solving the equation of motion for $V_t$ starting from (or near) the false vacuum. For pseudo-bounces in particular, the correct boundary conditions at the end point of the tunneling come out automatically (while they seem contrived in the Euclidean formulation). For bubbles of nothing, one gets naturally divergent $V_t$'s with finite action (interestingly, one such example was found in \cite{EFH} before the BoN interpretation was clear) which correspond
precisely to BoN solutions. Finally, for both BoN and AdS maxima decays, while the Euclidean formulation requires boundary terms on top of the usual 4d Euclidean action, the naive $S[V_t]$ reproduces the correct results without needing any adjustment.

Clearly, the $V_t$ formulation, which was first derived starting from the Euclidean formalism, has no claim to being more fundamental. As it stands, it is just a mathematical reformulation which happens to have some convenient properties, quite a few of them surprising. Is there some deep physical reason behind this behaviour? Is there some more fundamental way of understanding vacuum decay that would lead directly to the $V_t$ formulation? Coming full circle, one is precisely in the situation Wigner described in his article \cite{Wigner}: {\em "It is just this uncanny usefulness of mathematical concepts that raises the question of the uniqueness of our physical theories"}.

\end{document}